\documentclass[aps,prl,twocolumn,preprintnumbers,superscriptaddress]{revtex4-2}
\usepackage{graphicx}
\usepackage{physics}
\usepackage{amsmath}
\usepackage{amssymb}
\usepackage{bm}
\usepackage{color}
\definecolor{LinkColor}{rgb}{0.0,0.0,1}
\usepackage{hyperref}
\hypersetup{
        colorlinks=true,
        citecolor=LinkColor,
        linkcolor=LinkColor,
        urlcolor=LinkColor,
}

\begin{document}

\title{Topological Chiral Superconductivity from Doping Correlated \\Charge Transfer Insulator with Orbital Magnetism}

\author{Wen Sun}
\email{physunwen@smail.nju.edu.cn}
\affiliation{Great Bay Institute for Advanced Study, Great Bay University, Dongguan 523000, China}
\affiliation{School of Physical Sciences, Great Bay University, Dongguan 523000, China}
\affiliation{Hefei National Research Center for Physical Sciences at the Microscale and School of Physical Sciences,
University of Science and Technology of China, Hefei, Anhui 230026, China}

\author{Yuan Yang}
\affiliation{School of Physical Sciences, Great Bay University, Dongguan 523000, China}
\author{Shou-Shu Gong}
\email{shoushu.gong@gbu.edu.cn}
\affiliation{School of Physical Sciences, Great Bay University, Dongguan 523000, China}
\author{Wen Huang}
\email{huangwen@gbu.edu.cn}
\affiliation{School of Physical Sciences, Great Bay University, Dongguan 523000, China}

\date{\today}

\begin{abstract}
Topological chiral superconductivity has been extensively studied from the perspective of electronic structures in momentum space, while its emergence from real-space processes in strongly interacting systems remains largely unexplored. Here, we study a doped charge-transfer insulator in a strongly repulsive spinless-fermion model with three sublattices and $2\pi$ flux per unit cell. Combining perturbative analysis, exact diagonalization, and density matrix renormalization group simulations, we establish a topological superconducting ground state with $p+ip$ symmetry. In the large charge-transfer gap limit, chiral charge-transfer excitations generate a correlated-hopping process that provides a microscopic mechanism for electron pairing and its chiral structure. The chiral superconductivity persists into the smaller-gap regime, where chiral excitonic two-particle bound state can emerge. Our results establish a concrete real-space route to topological chiral superconductivity upon doping strongly interacting charge-transfer insulators with orbital magnetism.
\end{abstract}

\maketitle

\textit{Introduction.}---
Topological chiral superconductors can host non-Abelian Majorana zero modes in their vortex cores, with potential applications in fault-tolerant quantum computation \cite{Volovik_1999,PhysRevB.61.10267,PhysRevLett.86.268,RevModPhys.80.1083,Alicea_2012}. Recent experiments have revealed chiral superconductivity (SC) characterized by the coexistence of SC and orbital magnetism, in tunable 2D van der Waals systems such as rhombohedral multilayer graphene \cite{han2025Signatures,sheekey2026visualizingorbitalmagnetismelectron,dutta2026reconfigurablechiralsuperconductivity,kalantre2026fermiologycandidatechiralsuperconductor,zheng2026chiralsuperconductorscompetingstates} and twisted MoTe$_2$ \cite{xuSignaturesUnconventionalSuperconductivity2025}. Several microscopic mechanisms, including Kohn-Luttinger and anyon SC, have been proposed to account for chiral SC and its competition with other phases in these systems \cite{ghazaryan2021Unconventional,geierChiralTopologicalSuperconductivity2025,parra-martinez2025Band,Guerci2025,xu2025Chiral,yang2025Topological,Qin_2026,guerci2026topologicalsuperconductivityemergentvortex,li2025berrytrashcanshortrange,PhysRevB.111.174523,divic2025Anyon,pichler2026Microscopic,shi2025Doping,nosov2026Anyon,chen2026Topological,kuhlenkamp2025robustsuperconductivitydopingchiral,wangTopologicalSuperconductivityAbelian2026,fanHiddenWeakpairingSuperconductivity2026,PhysRevB.111.014508,f76m-jyn5}. While existing studies have largely focused on momentum-space properties of the normal state, where features such as flat band dispersion and quantum geometry can play an important role \cite{han2025exactmodelschiralflatband,gt8h-czf3,pgqh-wm5v,MayMann2026,1tg5-qhtf,yw2c-qyj1,PhysRevLett.134.176001,sw7f-4x94,f43d-414n,PhysRevResearch.6.043240,hu2026quantumgeometryinducedsuperconductivitynearfractional}, topological chiral SC in lattice systems beyond flat bands, particularly electronic mechanisms rooted in real-space processes, remains rarely explored.

Interestingly, charge-transfer insulators provide a setting for exploring such real-space mechanisms in multiband system. Upon doping, low-density charge carriers can pair and condense even in the presence of strong repulsion through correlated hopping mediated by charge-transfer excitations \cite{PhysRevB.102.235423,Cr_pel_2021}. This mechanism has been generalized and applied to various systems, leading to unconventional pairing symmetries \cite{doi:10.1073/pnas.2117735119,Cr_pel_2023,PhysRevB.106.L180502,PhysRevB.105.094506,PhysRevResearch.5.L012009,PhysRevB.111.125104,glxl-l925}. Whether it can also give rise to a topological chiral superconductor with an odd Bogoliubov--de Gennes (BdG) Chern number is still an open question. In the presence of orbital magnetism, the charge-transfer hopping process can acquire nontrivial phases, allowing the normal-state chirality to influence the superconducting pairing. These considerations raise the possibility of realizing topological chiral SC by doping a charge-transfer insulator with orbital magnetism.

In this work, we study a strongly repulsive spinless fermion model with three sublattices and a $2\pi$ flux per unit cell. At filling $n=2$, the system forms a charge-transfer insulator. 
Upon electron doping, density matrix renormalization group (DMRG) calculations reveal dominant, algebraically decaying pairing correlations with $p+ip$ symmetry over a broad range of charge-transfer gaps. In the large-gap limit, virtual particle-hole excitations across the charge-transfer gap generate a chiral correlated hopping term in the effective model. Exact diagonalization of this model further establishes the topological nature of the superconducting state in this regime. At smaller charge-transfer gaps, chiral excitonic two-particle bound states emerge and may provide an additional mechanism for electron pairing. Our results identify a microscopic route to topological chiral SC in a strongly interacting charge-transfer system.

\begin{figure}
    \centering
 \includegraphics[width=0.49\columnwidth]{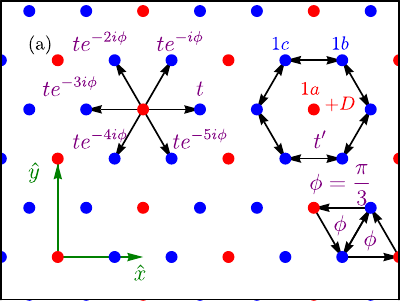}
    \hfill
        \includegraphics[width=0.49\columnwidth]{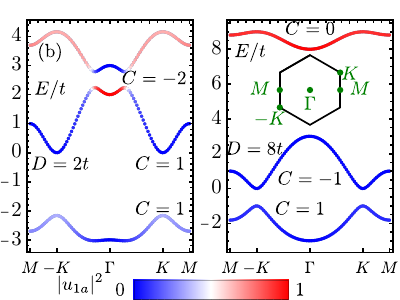}
        \caption{(a) Illustration of the spinless fermion model on a triangular lattice with $\sqrt{3}\times\sqrt{3}$-modulation. The $1a$ sites are colored red, while the $1b$ and $1c$ sites are colored blue. A positive on-site potential $D$ is applied to the $1a$ sites. The independent complex-hopping-amplitudes are indicated by arrows. Each triangular plaquette encloses a magnetic flux $\phi$, $\phi=\frac{\pi}{3}$. (b) Noninteracting band structures at different values of the potential $D$ for $t=t'$. The Chern number of each band are indicated. The band dispersions are color-coded according to the weight of the $1a$-component in the corresponding Bloch wavefunction.}
        \label{model}
\end{figure}

\textit{Model.}---
We consider a model of spinless fermions on a triangular lattice, threaded by a $\pi/3$-flux through each triangular plaquette. A positive on-site potential $D$ is applied to one sublattice of a $\sqrt{3}\times \sqrt{3}$ superlattice, such that the total flux through each superlattice unit cell is $2\pi$. The fermions experience a nearest-neighbor density-density repulsion $V$. The Hamiltonian is
\begin{equation}
        \label{eq:Hamiltonian}
        H=-\sum_{\langle ij\rangle} (t_{ij}c^\dagger_i c_j + h.c.)+ D\sum_{i\in1a}n_i+V\sum_{\langle ij\rangle}n_in_j.
\end{equation}
As shown in Fig.~\ref{model}(a), the $1a$ sites with potential $D$ possess $C_{6z}$ symmetry and form a triangular superlattice, while the $1b$ and $1c$ sites possess $C_{3z}$ symmetry and form a honeycomb superlattice. The hopping processes involving the $1a$ sites exhibit a chiral $p$-wave phase winding, as indicated by the complex hopping amplitudes in Fig.~\ref{model}(a). Equivalently, this model can be viewed as a system with $s$-orbitals on the $1b$ and $1c$ sites, and $(p-ip)$-orbitals on the $1a$ sites. For simplicity, we use {\it lattice} to denote the superlattice in the following discussion.

For two electrons per unit cell, the noninteracting system can realize a $C=2$ Chern insulator in the smaller-$D$ regime. As shown in Fig.~\ref{model}(b), increasing $D$ drives the system into a topologically trivial insulating state, with charges predominantly localized on the $1b$ and $1c$ sites. Upon electron doping, the doped carriers enter the $1a$ sites, while the parent state remains a charge-transfer insulator. This insulating state is adiabatically connected to an atomic insulator, which is the ground state in the classical $V\to\infty$ limit. We focus on electron doping of this insulating state with filling $n=2+x$ ($x>0$).

\begin{figure}
    \centering
 \includegraphics[width=0.49\columnwidth]{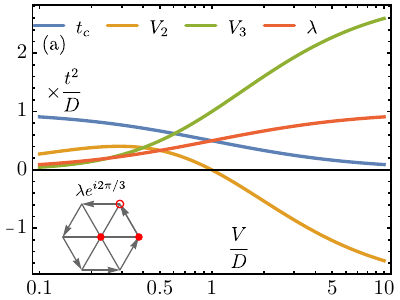}
    \hfill
        \includegraphics[width=0.49\columnwidth]{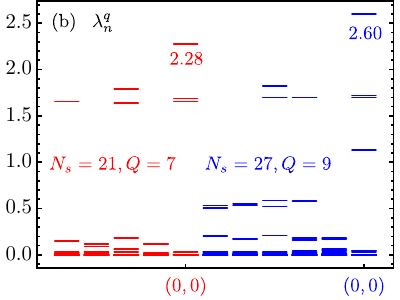}
    \vspace{2mm}
        \includegraphics[width=0.49\columnwidth]{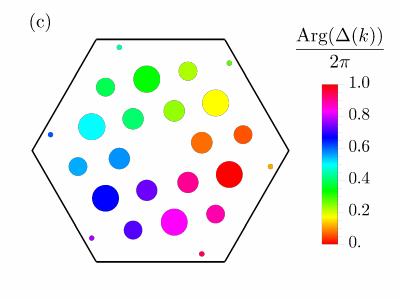}
        \hfill
        \includegraphics[width=0.49\columnwidth]{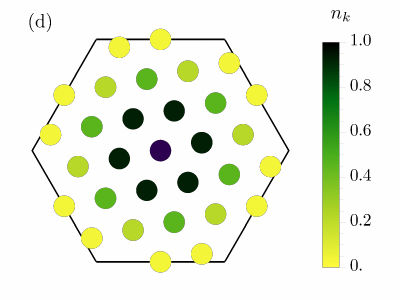}
        \vspace{2mm}
        \includegraphics[width=0.32\columnwidth]{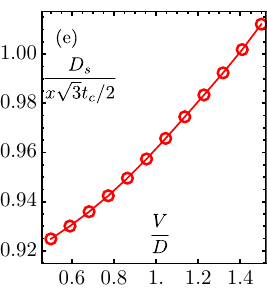}
        \hfill
        \includegraphics[width=0.32\columnwidth]{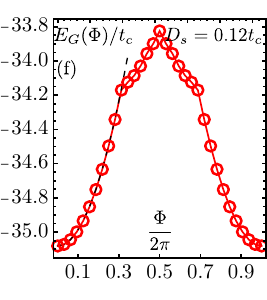}
        \hfill
        \includegraphics[width=0.32\columnwidth]{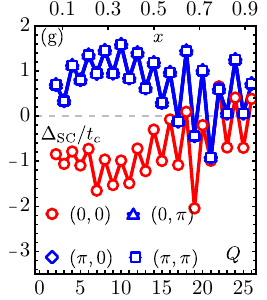}
        \caption{Exact diagonalization study of the effective single-band model on the triangular lattice. (a) Dependence of the effective Hamiltonian parameters on the bare repulsion $V$. The inset illustrates the chiral correlated-hopping process with amplitude $\lambda e^{i2\pi/3}$. (b) Eigenvalue spectrum of the two-particle density matrix for the $21$- and $27$-site systems at $1/3$ filling factor, shown in red and blue colors, respectively. (c) and (d) are pairing eigenfunction $\Delta(k)$ and momentum-space occupation number $n_k$, respectively, obtained from the ground state of $11$ electrons on $28$ sites. (e) Dependence of charge stiffness $D_s$ for the $49$-site system at $x=1/7$ filling on the bare repulsion.  (f) Variation of the ground-state energy $E_G(\Phi)$ with threaded magnetic flux $\Phi$ for $7$ electrons on $49$ sites. (g) Superconducting gap $\Delta_{\rm SC}$ as a function of particle number $Q$ (filling factor $x$) for different flux configurations on a $28$-site torus.}
        \label{singlebandmodel}
\end{figure}

\textit{Chiral correlated hopping in single-band model.}---
We first consider the large charge-transfer gap limit, $D\gg t$. In this regime, the doped carriers on the $1a$ sites are effectively described by a single-band model on the triangular lattice. Treating the hopping $t$ and $t'$ perturbatively, the low-energy charge carriers acquire effective hoppings and interactions through virtual particle-hole excitations across the charge-transfer gap. Applying a Schrieffer-Wolff transformation, we obtain the effective Hamiltonian
\begin{equation}
        \begin{split}
        H_{\text{eff}}&=-t_c\sum_{\langle ij\rangle }\left(c_i^\dagger c_j+h.c.\right)\\
         &+ V_2\sum_{\langle ij\rangle}n_in_j+V_3\sum_{ijk\in\triangle/\triangledown}n_in_jn_k\\
         &+\lambda\sum_{ijk\in\triangle/\triangledown}P_{\text{cyc}}\left( n_ic_k^\dagger c_je^{i\phi_{kj}} + h.c. \right),
        \end{split}
\end{equation}
where $P_{\text{cyc}}$ denotes the sum over three cyclic permutations of indices $i,j$ and $k$.
Figure~\ref{singlebandmodel}(a) shows the dependence of the effective model parameters on the interaction strength $V/D$ (see Supplementary Materials (SM)~\cite{supp} for derivations). In the noninteracting limit, the positive effective hopping $t_c$ gives rise to a single Fermi surface at low doping, in contrast to the two-valley structure found in the models of Refs.~\cite{Cr_pel_2021,PhysRevResearch.5.L012009}. The nearest-neighbor density interaction $V_2$ can become attractive for strong repulsion $V>D$, while the three-body interaction $V_3$ remains repulsive. Importantly, the virtual processes also generate a chiral correlated-hopping as illustrated in the inset of Fig.~\ref{singlebandmodel}(a), which describes a process where one carrier hops around another with a complex amplitude $\lambda e^{i2\pi/3}$, provided that the two carriers occupy neighboring sites. 

We first use exact diagonalization (ED) to study the ground state of the effective single-band model on finite-sized tori with at least $C_6$ symmetry (see SM~\cite{supp} for definitions). Below, we focus on the parameter regime with $V=D$, where both $\lambda$ and $V_3$ are nonzero. In SM~\cite{supp}, we independently tune these effective interactions to investigate their roles in pairing and demonstrate the importance of correlated hopping in promoting superconducting correlations. For each system size, we obtain the ground state $|G\rangle$ with conserved momentum quantum numbers, and analyze its two-particle reduced density matrix  
\begin{equation}
        \begin{split}
        &\rho^q_{k_1,k_2}=\langle G|c_{k_1+q}^\dagger c_{-k_1}^\dagger c_{-k_2}c_{k_2+q}|G\rangle,\\
        &\lambda^q_n\Delta_n(k_1)=\sum_{k_2}\rho^q_{k_1,k_2}\Delta_n(k_2).
        \end{split}
\end{equation} 
As shown in Fig.~\ref{singlebandmodel}(b) of the $21$- and $27$-site systems at $1/3$ filling, the largest eigenvalue $\lambda_n^q$ occurs in the zero-momentum $q = (0,0)$ pairing channel. Its growth with system size provides evidence for off-diagonal long-range order in large-size limit \cite{RevModPhys.34.694,guerci2026topologicalsuperconductivityemergentvortex,li2025berrytrashcanshortrange}. The corresponding dominant eigenvector $\Delta_n(k)$ characterizes the pairing function in momentum space and exhibits a $(p+ip)$-wave phase winding around the $\Gamma$ point of the Brillouin zone [Fig.~\ref{singlebandmodel}(c)]. Together with the single Fermi pocket structure shown in Fig.~\ref{singlebandmodel}(d), this pairing structure is consistent with a topological SC with Chern number $C=1$. 

The superconducting state is also characterized by its charge stiffness, which determines the Meissner effect. This quantity ($D_s$) can be obtained from the response of ground-state energy $E_G$ to a small magnetic flux $\Phi$ threaded through one hole of the torus \cite{Cr_pel_2021}, 
\begin{equation}
        D_s=\frac{L_1}{4L_2}\frac{d^2E_G(\Phi)}{d\Phi^2}\bigg|_{\Phi=0},
\end{equation}
where $L_1$ and $L_2$ are the linear system size in units of two basis vectors of a triangular lattice. As shown in Fig.~\ref{singlebandmodel}(e), its value is comparable to the charge stiffness of a BCS superconductor with a parabolic band dispersion over a broad range of interaction strength $V/D$. At low filling $x=1/7$, the later is $\sqrt{3}xt_c/2$ \cite{PhysRevX.9.031049}. Furthermore, we find that the ground-state energy is periodic in the threaded flux with periodicity $2\pi$ \cite{li2020fractionallittleparkseffectobserved} [Fig.~\ref{singlebandmodel}(f)], in contrast to the $\pi$-periodicity expected for a topologically trivial superconductor whose low-energy degrees of freedom are solely charge-$2e$ Cooper pairs \cite{Cr_pel_2021}.

We observe a cusp at $\Phi=\pi$ in Fig.~\ref{singlebandmodel}(f), indicating a reversal of the ground-state charge current across this flux. 
To further characterize the response to flux insertion, we examine the superconducting gap $\Delta_{\text{SC},Q}$ \cite{Cr_pel_2021} after inserting a $\pi$-flux through each of the two noncontractible holes of the torus,
\begin{equation}
        \Delta_{\text{SC},Q}=\frac{(-1)^Q}{2}\left(E_{Q+1}+E_{Q-1}-2E_{Q}\right),
\end{equation}
where $E_Q$ is the ground-state energy in the sector with $Q$ particles and the quantity in bracket gives the particle-hole excitation energy. The even-odd effect in $E_Q$ for a superconducting phase implies that  the corresponding $\Delta_{\text{SC}}$ has a fixed sign \cite{Guerci2025,Fu_2010}. As shown in Fig.~\ref{singlebandmodel}(g),  a finite gap with a consistent sign is found throughout the doping range where the Fermi surface consists of a single pocket, across all four flux configurations. At zero flux $(0,0)$, $\Delta_{\text{SC}}$ is negative, indicating an odd parity ground state. In contrast, the other three flux sectors have Fermi parity $+1$. The resulting fermion-parity switch under $\pi$-flux insertion provides a many-body signature of the nontrivial topology of the superconducting ground state \cite{PhysRevLett.113.267002,Ortiz_2016}.

\textit{Dominant $(p+ip)$-wave pairing correlations.}---
Encouraged by the ED evidence of topological SC in the effective single-band model, we use the DMRG~\cite{PhysRevLett.69.2863,RevModPhys.77.259,Fishman2022} method to simulate the three-band model defined in Eq.~\eqref{eq:Hamiltonian} on a YC cylinder geometry. The cylinder is periodic along the $y$-direction and open along the $x$-direction, with the corresponding unit vectors shown in Fig.~\ref{model}(a). The circumference and length of the cylinder are given by $L_y|\hat{y}|$ and $L_x|\hat{x}|$, respectively, with $L_y=4$ and $L_x=48$ for results presented in the main text; results for $L_y=3$ are provided in SM~\cite{supp}. For correlation functions, we fix the reference point $r_0$ at the $\frac{L_x}{4}$-th column of the cylinder to minimize boundary effects. The DMRG simulation preserves the charge $U(1)$ symmetry, with bond dimensions up to $6000$. The calculations achieve truncation errors of order $10^{-8}$ to $10^{-9}$ (see SM~\cite{supp} for convergence check). 

\begin{figure}
    \centering
 \includegraphics[width=0.49\columnwidth]{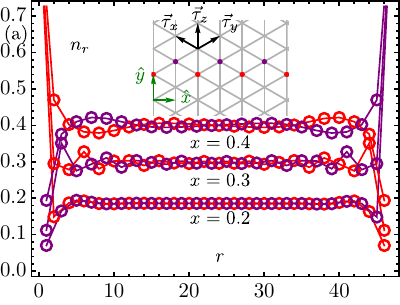}
    \hfill
        \includegraphics[width=0.49\columnwidth]{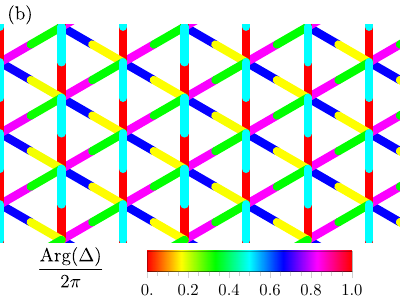}
    \vspace{2mm}
        \includegraphics[width=0.49\columnwidth]{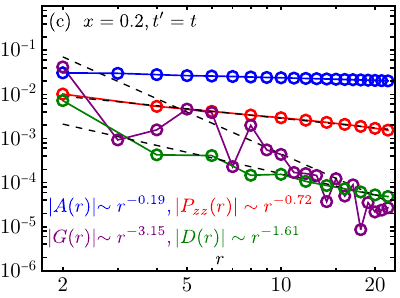}
        \hfill
        \includegraphics[width=0.49\columnwidth]{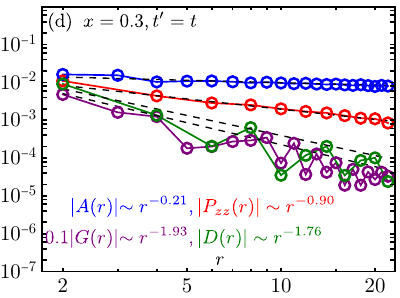}
        \vspace{2mm}
        \includegraphics[width=0.49\columnwidth]{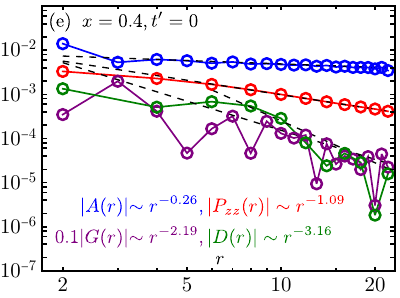}
        \hfill
        \includegraphics[width=0.49\columnwidth]{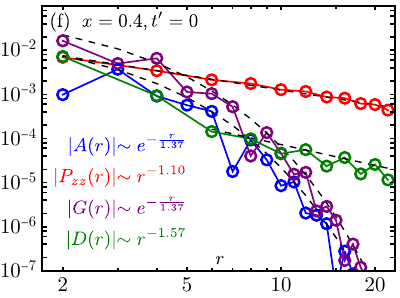}
        \caption{DMRG results of the three-band model in the large charge-transfer gap limit with $D = V = 8t$. (a) Charge distribution on two rows of $1a$ sites (red and purple of the inset for $1a$ sites) for doping levels $x = 0.2 - 0.4$. (b) The phase distribution of superconducting order parameter on nearest-neighbor bonds. On the same bond, two kind of colors illustrate the phases with $\pi$ difference due to the fermion statistics. (c-f) Double-logarithmic plots of correlation functions for different doping levels and hopping amplitudes ($t'$). A $\pi$-flux is inserted for the Hamiltonian in (f).}
        \label{largeD}
\end{figure}

We first consider $D=V=8t$, where the effective single-band description is expected to be valid. We focus on the observables associated with the $1a$ sites to probe the low-energy properties of the model. As shown in Fig.~\ref{largeD}(a), the $1a$-site charge density remains uniform in the bulk without static charge order for doping levels $x = 0.2-0.4$.  The superconducting tendency is detected by measuring the pairing correlation functions
\begin{equation}
        \label{eq:paircorrelation}
        P_{\alpha\beta}(r)=\langle c_{r_0+r}c_{r_0+r+\tau_\alpha}c_{r_0+\tau_\beta}^\dagger c_{r_0}^\dagger \rangle,
\end{equation}
where $\vec{\tau}_{\alpha}(\alpha =x,y,z)$ denote the nearest-neighbor bond vectors [see the inset of Fig.~\ref{largeD}(a)]. 
The obtained pairing correlations are complex, and the phase distribution of the bond pairs can be extracted by fixing one bond $\beta=z$ in $P_{\alpha\beta}(r)$. As shown in Fig.~\ref{largeD}(b), the phase pattern is translationally invariant in the bulk and exhibits a $(p+ip)$-phase winding around each $1a$ site.
The magnitudes of pairing correlations $|P_{\alpha\beta}(r)|$ exhibit a power-law decay with distance, $|P_{\alpha\beta}(r)|\sim r^{-K_{\alpha\beta}}$. As shown in Fig.~\ref{largeD}(c-f), $|P_{zz}(r)|$ decays slowly with $K_{zz} \lesssim 1$. The other components $|P_{\alpha\beta}(r)|$ decay with the same magnitude and exponent $K_\text{SC}$ (see SM~\cite{supp}), indicating that pairing correlations are isotropic despite the quasi-1D geometry. We also examine the charge density correlation $D(r)= \langle n_{r_0+r}n_{r_0}\rangle -\langle n_{r_0+r}\rangle \langle n_{r_0}\rangle$,
which decays algebraically but with an exponent larger than $K_{\rm SC}$. This indicates that superconducting correlatins constitute the leading ordering tendency at the studied doping levels and hopping parameters. With increasing doping, pairing correlations gradually weaken [Fig.~\ref{largeD}(c-e)], consistent with the closing of the superconducting gap at larger doping levels $(x>0.5)$ observed in the ED calculation.

\textit{1D topology in single-particle sector.}---
For a single wire of spinless fermions, theory and numerics have established that, in the absence of charge order, the Luttinger liquid phase is gapless to charge-$1e$ excitations, whereas the strongly paired phase is gapped to such excitations \cite{Kane2017,physrevb.98.214501,Ruhman2017}. Motivated by this \textit{one-dimensional classification}, we examine the single-particle correlations between $1a$ sites  $G(r)=\langle c_{r_0}^\dagger c_{r+r_0}\rangle$ in our quasi-1D system.
Under the periodic boundary conditions along the $y$-direction [Fig.~\ref{largeD}(c-e)], the single-particle excitation appears to be gapless, manifest as the power-law decay of $|G(r)|$. %
The system is in a gapless Luttinger liquid regime with dominant chiral pairing correlations. 

At the mean-field level, the single-particle sector of spinless fermions in 1D can be mapped onto the topological and trivial phases of the Kitaev chain \cite{Kitaev_2001}. Following the diagnostic proposed in Ref.~\cite{YixinZhang}, this topological structure persists in the presence of 1D superconducting phase fluctuations and can be detected through the nonlocal string operator
\begin{equation}
A(r)=\langle c_{r_0}^\dagger(-1)^{F_{r_0,r}} c_{r_0+r}\rangle, F_{r_0,r}=\sum_{r_0<i<r_0+r}n_i,
\end{equation}
where $r_0$ and $r_0 + r$ are $1a$ sites, and the sum in $F_{r_0,r}$ runs over all single-particle modes between them according to the linear ordering of the matrix-product state, including the intervening $1b$ and $1c$ sites. In the Luttinger liquid regime, the magnitude $|A(r)|$ exhibits a power-law decay with distance, as shown in Fig.~\ref{largeD}(c-e). Importantly, the decay exponent extracted from $|A(r)|\sim r^{-K_s}$ satisfies $K_s\approx\frac{1}{4}K_{\text{SC}}$, in agreement with the relation proposed in Ref.~\cite{YixinZhang}. For the system threaded by a $\pi$-flux through the hollow of the cylinder, as shown in Fig.~\ref{largeD}(f), $|A(r)|$ exhibits exponential decay and the single-particle correlation also decays rapidly. Meanwhile, superconducting correlations remain the leading ordering tendency, consistent with a strongly paired regime. Interestingly, we observe nonlocal single-particle correlations between the two edges in the Luttinger-liquid regime, potentially reflecting the underlying bulk topology in the single-particle sector (see SM~\cite{supp}).

\textit{Chiral excitonic Cooper pair.}---
Since the correlated insulator in the large charge-transfer gap limit $(D\gg t)$ is adiabatically connected to the generalized Wigner crystal stabilized by strong repulsion $(V\gg t)$ and weakly pinned by $D$, it is natural to ask whether the doping-induced SC persists as $D$ is reduced in the strongly coupled regime. Before addressing this question, we examine the few-body excitations of the doped insulator in the strong-coupling limit. Remarkably, the two-particle state can lower its energy by coherently hybridizing with excitonic configurations, forming a chiral excitonic Cooper pair \cite{guerci-2026}.

\begin{figure}
    \centering
 \includegraphics[width=0.49\columnwidth]{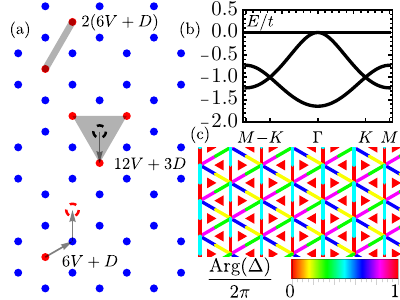}
    \hfill
        \includegraphics[width=0.49\columnwidth]{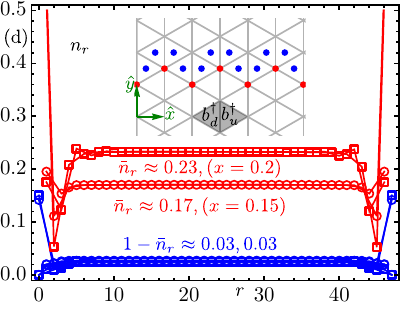}
 \vspace{2mm}
        \includegraphics[width=0.49\columnwidth]{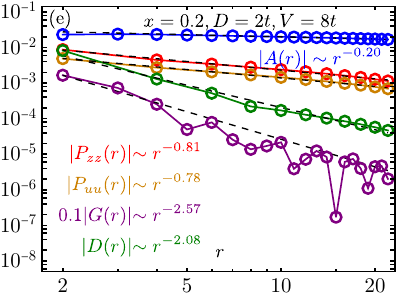}
        \hfill
        \includegraphics[width=0.49\columnwidth]{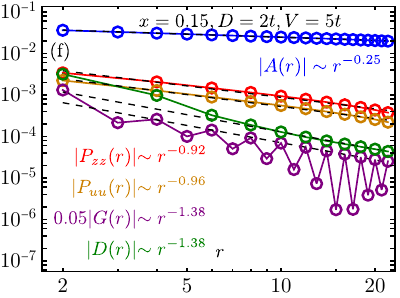}       
        \caption{Excitonic Cooper pair and DMRG results for the three-band model in the moderate charge-transfer-gap regime with  $D=2t,t'=t$. (a) Illustration of the hopping and excitonic hybridization processes for one- and two-particle doped states. The charge-transfer-insulator background is shown in blue, and the doped carriers in red. (b) Lowest three energy levels of the two-particle bound states, with energies measured relative to two spatially separated doped carriers. (c) Phase distribution of the bond-pair operators on the triangle edge and the trimer operators at the triangle centers. (d) Charge density on the $1a$ (red) and $1b/c$ (blue) sites. (e,f) Double-logarithmic plots of the corresponding correlation functions for different doping levels and interaction strengths. Here, $P_{uu}(r)=\langle b_{r_0+r,u}b_{r_0,u}^\dagger\rangle$.}
        \label{excitoniccooper}
\end{figure}

A single doped electron costs an energy $(6V+D)$ on a $1a$ site and is immobile in the strong-coupling limit. In contrast, a neighboring pair of doped electrons can coherently tunnel into trimer configurations, in which three electrons bind with a hole in the insulating background without paying an additional energy of order $V$, as illustrated in Fig.~\ref{excitoniccooper}(a). The lowest two-particle state occurs at the $\Gamma$ point of the Brillouin zone [Fig.~\ref{excitoniccooper}(b)]. Its binding energy $(E_b<0)$ and the wavefunction are given by
\begin{equation}
        \begin{split}
        E_b&=\frac{1}{2}\left(D-\sqrt{D^2+24t^2}\right),\\
        |\psi\rangle &= \sum_r\left(e^{i2\pi/3}b_{r,y}^\dagger +e^{i\pi/3}b_{r,z}^\dagger +b_{r,x}^\dagger \right)\\
        &+\sum_r\frac{6te^{i\pi/3}}{D+\sqrt{D^2+24t^2}}\left(b_{r,u}^\dagger +b_{r,d}^\dagger \right),
        \end{split}
        \label{eq:excitoniccooper}
\end{equation}
where $b_{r,\alpha}^\dagger=c_r^\dagger c_{r+\tau_\alpha}^\dagger,b_{r,u}^\dagger =c_r^\dagger c_{r+\tau_y}^\dagger c_{r+\tau_z}^\dagger c_{r,b}, b_{r,d}^\dagger =c_r^\dagger c_{r+\tau_z}^\dagger c_{r+\tau_x}^\dagger c_{r,c}$ and $c_{r,b}\equiv c_{r+\frac{\tau_y+\tau_z}{3}},c_{r,c}\equiv c_{r+\frac{\tau_z+\tau_x}{3}}$ with the labels defined in the insets of Fig.~\ref{largeD}(a) and Fig.~\ref{excitoniccooper}(d).
The trimer weight and the binding energy $|E_b|$ are both maximized at $D=0$, where the undoped system spontaneously forms the
generalized Wigner crystal. In this limit, however, the doped system is unstable toward phase separation \cite{wessel2005Supersolid,Kourtis_2018}, suggesting that a finite charge-transfer potential $D$ is necessary to stabilize SC \cite{guerci-2026}, which we investigate below using DMRG.

As shown in Fig.~\ref{excitoniccooper}(d) for $D=2t$, the charge distribution remains uniform in the bulk, with a finite charge transfer from the $1b$ and $1c$ sites to the $1a$ sites. To elucidate the excitonic nature of the Cooper pairing, we measure the correlations of the trimer and bond-pair operators defined in Eq.~\eqref{eq:excitoniccooper} and Eq.~\eqref{eq:paircorrelation}, respectively, as well as their cross-correlations. Following the procedure used for Fig.~\ref{largeD}(b), we extract the phase structure of the corresponding pairing correlations [Fig.~\ref{excitoniccooper}(c)]. The bond-pair correlations exhibit a $(p+ip)$-wave phase winding around each $1a$ site, while the trimer correlations $P_{\alpha \beta}(r)=\langle b_{r_0+r,\alpha}b_{r_0,\beta}^\dagger\rangle$ ($\alpha,\beta = u,d$) have a uniform phase, in agreement with the chiral excitonic Cooper-pair wavefunction in Eq.~\eqref{eq:excitoniccooper}. In this case, complementary Hartree-Fock, random-phase-approximation, and BCS mean-field calculations also identify a $p+ip$ superconducting ground state with BdG Chern number $C=1$ (see SM \cite{supp}). The bond-pair and trimer correlations have comparable magnitudes and both decay algebraically, as shown in Fig.~\ref{excitoniccooper}(e) and (f), with exponents smaller than that of the charge density correlations. The single-particle Green's function and nonlocal string correlator likewise exhibit power-law decay, with the later satisfying $K_s\approx\frac{1}{4}K_{\text{SC}}$. Thus, the excitonic Cooper-pair regime exhibits the same gapless single-particle and topological string signatures as the large-gap regime, providing evidence for chiral SC beyond the regime described by the effective single-band theory.

\textit{Summary and discussion.}---
We have demonstrated a real-space mechanism for topological chiral SC by doping a strongly interacting charge-transfer insulator with orbital magnetism. In the large charge-transfer-gap limit, virtual charge-transfer processes generate chiral correlated hopping in the effective model, producing a $(p+ip)$-superconducting ground state with $C=1$. The SC is not confined to the perturbative regime. At smaller charge-transfer gaps, the Cooper pairs acquire chiral excitonic character through coherent hybridization with the particle-hole excitations of the insulating background. Broadly, our results suggest that charge-transfer processes in strongly correlated multiband lattice systems provide a useful setting for engineering topological chiral SC.

The experimental realization of our mechanism appears promising in moiré systems, where emergent magnetic fields have been extensively explored to engineer topological flat bands \cite{PhysRevLett.122.086402,ledwith2023vortexability,ledwith2022family,wang2021exact,PhysRevLett.132.096602,doi:10.1126/sciadv.abn1401,PhysRevLett.133.206601,PhysRevB.111.L241104,Liu_2022,z4g8-2kwj,yang2025Fractional,Schlawin2022,PhysRevLett.132.166901,Lin2023,fcy2-2py5,PhysRevX.15.011004,n824-tw3h,liu2026idealtopologicalflatbands,dong2022diracelectronperiodicmagnetic}. Strong intra-unit-cell potential modulation, which favors charge localization and drives the parent state into a topologically trivial charge-transfer insulator, has received comparatively less attention in this context \cite{Zhang2024,Gu2024,He2024,liu2025fractionalcherninsulatorstopological,Ding2024,h2k9-7v61}. Our results suggest that combining such charge inhomogeneity with orbital magnetism provides a complementary route to topological SC upon doping, without requiring a topologically nontrivial parent band structure.

\textit{Note added.}---
In the final preparation of this manuscript, Ref.~\cite{gallegos2026repulsiondrivenpi} appeared on arXiv, reporting DMRG evidence for repulsion-driven topological chiral superconductivity in the doped Qi–Wu–Zhang model. Their work emphasizes the role of the nontrivial quantum geometry of the parent bands, whereas our work identifies a real-space mechanism based on charge-transfer processes in a strongly interacting charge-transfer insulator.

\textit{Acknowledgements.}--- 
We acknowledge the organizers of M2S 2026, where preliminary results of this work were presented. We thank Xianbiao Shi, Hong Yao and Yifan Jiang for helpful discussions.
W. S. and W. H. were supported by the NSFC under Grant No.~12374042 and the Guangdong Provincial Quantum Science Strategic Initiative (GDZX2601007). 
Y. Y. was supported by the Guangdong Basic and Applied Basic Research Foundation (No.~2025A1515110087).
S. S. G. was supported by the NSFC (No. 12534009 and No. 12274014), the Guangdong Provincial Quantum Science Strategic Initiative (No. GDZX2501006), the Special Project in Key Areas for Universities in Guangdong Province (No. 2023ZDZX3054), and the Dongguan Key Laboratory of Artificial Intelligence Design for Advanced Materials.
The computational resources were supported by the SongShan Lake HPC Center at Great Bay University. 
We acknowledge the use of ChatGPT (OpenAI) for assistance with code optimization and implementation. All AI-generated suggestions and code were reviewed and verified by the authors, who take full responsibility for the numerical implementation and the results reported in this work.

\textit{Data availability.}---The data are available from the authors upon reasonable request.

\bibliography{extracted.bib}
\appendix
\widetext
\newpage
\begin{center}
        \textbf{\large Supplemental Materials for: Topological Superconductor from \\Doping Correlated Charge Transfer Insulator with Orbital Magnetism}
\end{center}
\vspace{1mm}
\renewcommand\thefigure{\thesection S\arabic{figure}}
\renewcommand\theequation{\thesection S\arabic{equation}}
\setcounter{figure}{0}
\setcounter{equation}{0}

\tableofcontents

\section{Effective single-band model}
Here we derive an effective single-band model on the triangular lattice from the $\frac{\pi}{3}$-hofstadter model at filling $n=2+x$ $(x>0)$ in the small hopping limit. The full three-band Hamiltonian is 
\begin{equation}
        H=-\sum_{\langle ij\rangle}\left(t_{ij}c_i^\dagger c_j +\text{H.c.}\right)+V\sum_{\langle ij\rangle} n_in_j + D\sum_{i\in 1a} n_i
\end{equation}
At filling $n=2$, the system forms an atomic insulator with charges localized on the $1b$ and $1c$ sites, provided that the hopping amplitude $|t_{ij}|$ are small compared to the charge-transfer gap $D$ $(D>0)$. Upon electron doping, the additional electrons occupy the $1a$ sites, which form a triangular lattice. In the strong-coupling limit considered below, the direct hopping $t'$ between the $1b$ and $1c$ sites acts only within the inert insulating background and therefore does not contribute to the second-order effective dynamics within the low-energy subspace of doped $1a$ electrons.

The hopping $t$ can generate an effective dynamics for the doped electrons by coupling the low-energy subspace to high-energy particle-hole fluctuations of the atomic insulator. These virtual processes can be systematically incorporated using a Schrieffer-Wolff transformation. We decompose the three-band Hamiltonian as 
\begin{equation}
        \begin{split}
        H&=T+U+T'\\
        T&=-\sum_{i\in 1a,j\in1b/1c}\left(t_{ij}c_i^\dagger c_j +\text{H.c.}\right)\\
        T'&=-\sum_{\langle ij\rangle \in1b\&1c}\left(t_{ij}c_i^\dagger c_j+\text{H.c.}\right)\\
        U&=V\sum_{\langle ij\rangle} n_in_j + D\sum_{i\in 1a} n_i.
        \end{split}
\end{equation}
Here, $U$ is the diagonal in the occupation-number basis, while $T$ connects the low-energy and high-energy subspaces. The hoppings $T'$ acts within the $1b/1c$ sites and has no matrix elements between these subspaces. We introduce an anti-Hermitian generator $S$ and perform the unitary transformation $e^SHe^{-S}$, choosing $S$ order by order to eliminate the off-diagonal matrix elements. To second order in $\frac{t}{D}$, 
\begin{equation}
        \begin{split}
        e^SHe^{-S}&=U+T'+T+[S,U]+[S,T]+[S,T']+\frac{1}{2}[S,[S,U]]+\cdots\\
                T&+[S,U]=0\\
                H_{\text{eff}}&=P(T'+U+\frac{1}{2}[S,T]+[S,T'])P.
        \end{split}
\end{equation}
Here, $P$ projects onto the low-energy subspace $\mathcal{H}_l$, in which all $1b$ and $1c$ sites are occupied, while $Q=1-P$ projects onto the high-energy subspace $\mathcal{H}_h$. Since the $1b$ and $1c$ sites are fully occupied within $\mathcal{H}_l$, we have $PT'P=0,P[S,T']P=0$, and hence the leading nontrivial contribution to the effective dynamics is generated by the second-order processes involving $T$. The effective Hamiltonian for a fixed doping level is expanded as
\begin{equation}
        H_{\text{eff}}=\frac{1}{2}\sum_{f,i\in\mathcal{H}_l}\sum_{m\in\mathcal{H}_h}\left(|f\rangle \langle f|S|m\rangle \langle m |T|i\rangle \langle i|-|f\rangle \langle f|T|m\rangle \langle m |S|i\rangle \langle i|\right)
\end{equation}
The matrix elements of the generator are determined by
\begin{equation}
        \langle f|S|i\rangle =\frac{\langle f|T|i\rangle}{E_f-E_i}
\end{equation}
where $E_i$ and $E_f$ are the eigenenergies of $U$ in the corresponding states.
The two terms in the second-order effective Hamiltonian are related by Hermitian conjugation. In particular,
\begin{equation}
        \begin{split}
        &\sum_{f,i\in\mathcal{H}_l}\sum_{m\in\mathcal{H}_h}|i\rangle \frac{\langle m|T|i\rangle^*\langle f|T|m\rangle^*}{E_f-E_m}\langle f|\\
        =&\sum_{f,i\in\mathcal{H}_l}\sum_{m\in\mathcal{H}_h}|i\rangle \frac{\langle i|T|m\rangle\langle m|T|f\rangle}{E_f-E_m}\langle f|\\
        =&-\sum_{f,i\in\mathcal{H}_l}\sum_{m\in\mathcal{H}_h}|f\rangle \frac{\langle f|T|m\rangle\langle m|T|i\rangle}{E_m-E_i}\langle i|
        \end{split}
\end{equation}
Therefore, the two contributions are Hermitian conjugates of each other, and it is sufficient to evaluate one of them; the other follows directly from Hermitian conjugation.
\begin{equation}
        \begin{split}
        PSQ& = -\sum_{R\in1b/1c}\sum_{R+\tau_i\in1a}\frac{t_{R,\tau_1}c_{R}^\dagger c_{R+\tau_1}}{-D+V(n_{R+\tau_2}+n_{R+\tau_3}-2)}\\
        H_{\text{eff}}&\ni\frac{1}{2}\sum_{R\in1b/1c}\sum_{\tau_2,\tau_2'}t_{R,\tau_2}t_{R,\tau_2'}^*c_{R}^\dagger c_{R+\tau_2}\frac{1}{-D+V(n_{R+\tau_1}+n_{R+\tau_3}-2)}c_{R+\tau_2'}^\dagger c_R
        \end{split}
\end{equation}
where $R+\tau_i,i=1,2,3$ are neighboring $1a$ sites of $R$. The second-order processes can be classified into two types. 
\begin{figure}
    \centering
    \includegraphics[width=0.24\columnwidth]{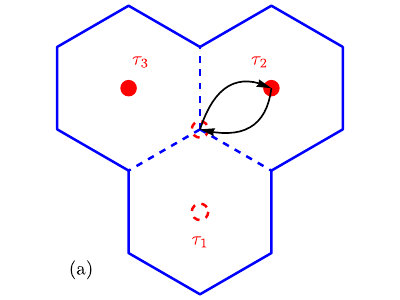}
    \includegraphics[width=0.24\columnwidth]{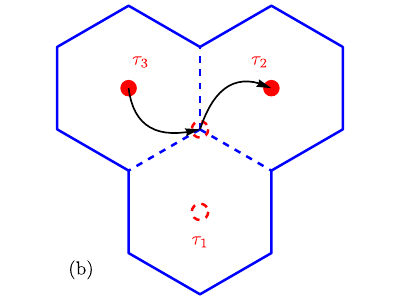}
     \caption{Second-order hopping processes for a doped charge carrier: (a) a particle hops to a neighboring site and returns to its initial site; (b) a particle hops between sites via an intermediate virtual state.}
    \label{secondorder}
\end{figure}
First, an electron in the atomic insulator virtually hops from a $1b$ or $1c$ site to a neighboring $1a$ site and subsequently returns, as illustrated in Fig.~\ref{secondorder}(a). The corresponding contribution is obtained by setting $\tau_2=\tau_2'$. Using the site labels of the effective triangular lattice, this contribution becomes
\begin{equation}
        t^2\sum_{1,2,3\in\bigtriangleup/\bigtriangledown}\left(\frac{1-n_1}{-D+V(n_2+n_3-2)}+\frac{1-n_2}{-D+V(n_3+n_1-2)}+\frac{1-n_3}{-D+V(n_1+n_2-2)}\right)
\end{equation}
Expanding this expression in the occupation numbers gives the two-body and three-body interactions,

\begin{equation}
        \begin{split}
       H_{\text{eff}}&\ni\sum_{\langle ij\rangle}V_2n_in_j +\sum_{1,2,3\in\bigtriangleup/\bigtriangledown}V_3n_1n_2n_3 \\
       V_2&=2t^2\left(\frac{4}{D+V}-\frac{1}{D}-\frac{3}{D+2V}\right)\\
       V_3&=3t^2\left(\frac{1}{D}+\frac{1}{D+2V}-\frac{2}{D+V}\right)
        \end{split}
\end{equation}
The second class of processes involves the virtual creation of a hole on a $1b$ or $1c$ site, which subsequently recombines with a doped electron on a neighboring $1a$ site, as shown in Fig.~\ref{secondorder}(b). For $\tau_2\neq\tau_3$, these processes generate effective hopping between neighboring sites of the triangular lattice:
\begin{equation}
        \frac{1}{2}\sum_{R\in1b/1c}\sum_{\tau_3\neq\tau_2}c_{R+\tau_3}\frac{t_{R,\tau_3}t_{R,\tau_2}^*}{-D+V(n_{R+\tau_1}-1)}c_{R+\tau_2}^\dagger+\text{H.c.}
\end{equation}
Decomposing the occupation-dependent denominator into density-independent and density-dependent parts yields the conventional hopping and chiral correlated-hopping terms,
\begin{equation}
        \begin{split}
        H_{\text{eff}}&\ni -t_c\sum_{\langle ij\rangle}\left(c_i^\dagger c_j+c_j^\dagger c_i\right)+\lambda\sum_R\sum_{j=1}^6\left(n_Rc_{R+a_{j+1}}^\dagger c_{R+a_j}e^{i2\pi/3}+n_Rc_{R+a_{j}}^\dagger c_{R+a_{j+1}}e^{-i2\pi/3}\right)\\
        t_c&=\frac{t^2}{D+V},
        \lambda = \frac{t^2}{D}-\frac{t^2}{D+V}
        \end{split}
\end{equation}
where $\vec{a}_j=R_6^j \vec{a}_0,\vec{a}_0=(1,0)$. 

To isolate the role of chiral correlated-hopping process, we consider the limit of $V\to\infty$. In this limit, the relevant two-body Hamiltonian reduces to
\begin{equation}
        \lambda \sum_R\sum_{j=1}^6\left( n_Rc_{R+a_{j+1}}^\dagger c_{R+a_j}e^{i2\pi/3}+n_Rc_{R+a_{j}}^\dagger c_{R+a_{j+1}}e^{-i2\pi/3}\right)+V_2\sum_{\langle ij\rangle}n_in_j +V_3\sum_{1,2,3\in \triangle,\triangledown}n_1n_2n_3
\end{equation}
with $\lambda=\frac{t^2}{D},V_2=-2\lambda,V_3=3\lambda$. When two doped particles occupy a nearest-neighbor bond, they can lower their energy through the combined effect of the correlated hopping and attractive density-interaction terms. In the $V\to\infty$ limit, isolated particles are immobile, while nearest-neighbor pairs can propagate through the correlated-hopping processes. There are three inequivalent nearest-neighbor bond configurations per triangular-lattice unit cell, which we choose as
\begin{equation}
        \begin{split}
                b_1^\dagger &=c_r^\dagger c_{r+a_1}^\dagger\\
                b_2^\dagger &=c_r^\dagger c_{r+a_2}^\dagger \\
                b_3^\dagger &= c_r^\dagger c_{r+a_2-a_1}^\dagger 
        \end{split}
\end{equation}
In the basis $(b_1,b_2,b_3)$, the two-particle Hamiltonian at momentum $k$ takes the form
\begin{equation}
\begin{pmatrix}
        V_2&\lambda e^{-i2\pi/3}\left(1+e^{ik(a_1-a_2)}\right)&-\lambda e^{i2\pi/3}\left(e^{ika_1}+e^{ik(a_1-a_2)}\right)\\
        \lambda e^{i2\pi/3}\left(1+e^{-ik(a_1-a_2)}\right)&V_2&\lambda e^{-i2\pi/3}\left(1+e^{ika_1}\right)\\
        -\lambda e^{-i2\pi/3}\left(e^{-ika_1}+e^{-ik(a_1-a_2)}\right)&\lambda e^{i2\pi/3}\left(1+e^{-ika_1}\right)&V_2
\end{pmatrix}
\end{equation}
The lowest-energy eigenstate occurs at the $\Gamma$ point and has energy
\begin{equation}
        E=-6\frac{t^2}{D},|\psi\rangle= \frac{1}{\sqrt{3N}}\sum_r\left(e^{-i\pi/3}b_{r,1}^\dagger +e^{-i2\pi/3}b_{r,2}^\dagger -b_{r,3}^\dagger \right)|0\rangle
\end{equation}

\section{Chiral Excitonic Cooper pair}
We consider the triangular lattice Hofstadter spinless fermion model with $\sqrt{3}\times\sqrt{3}$ modulated onsite potential. 
\begin{equation}
        H=-t\sum_{\langle ij\rangle} c_i^\dagger c_j e^{i\phi_{ij}}+\text{H.c.}+V\sum_{\langle ij\rangle }n_in_j + D\sum_{i\in 1a}n_i
\end{equation}
At a filling of two electrons per enlarged unit cell, the strong nearest-neighbor repulsion favors a honeycomb charge configuration, which is pinned by the onsite potential $D$ on the $1a$ sites. We consider electron doping of this insulating state. The doped particles preferentially occupy the $1a$ sites, which are located at the centers of the honeycomb plaquettes. 

In the strong coupling regime, it is useful to organize the Hilbert space according to the number of occupied nearest-neighbor bonds. We denote by $P_M$ the projector onto the subspace containing $M$ occupied bonds relative to the reference honeycomb configuration. 
We separate the Hamiltonian as
\begin{equation}
\begin{split}
        H&=H_0+H'\\
H_0&=V\sum_MP_MM+DN_a^M+T_{0,M}\\
H'&=\sum_{q\neq 0}T_{q,M}
\end{split}
\end{equation}
Here, $N_a^M$ denotes the number operator for particles occupying the $1a$ sites within the subspace $P_M$. $T_{q,M}$ is the projected hopping operator $P_{q+M}TP_M$. Here, $q$ denotes the change in the number of occupied bonds induced by a single hopping process. In general, $T_{0,M}$ does not commute with $N_{a}^M$, so the spectrum within a fixed-$M$ sector is not completely flat. 

We apply the Schrieffer-Wolff transformation to the above Hamiltonian. We introduce an anti-Hermitian generator $S=S^1+S^2+\cdots$, where $S^n$ is of order $\left(\frac{t}{V}\right)^n$. In the zeroth order, the Hamiltonian is just $V\sum_MP_MM + D\sum_MN_a^M $. The first order contribution is
\begin{equation}
        \sum_MT_{0,M}+\sum_{q\neq 0,M}T_{q,M}+[S^1,V\sum_MMP_M+D\sum_MN_a^M]
\end{equation}
We choose $S^1$ such that all matrix elements connecting different $M$ sectors vanish,
\begin{equation}
        \sum_{q\neq0,M}T_{q,M}+[S^1,V\sum_MMP_M+D\sum_MN_a^M]=0
\end{equation}
To second order, the transformed Hamiltonian contains
\begin{equation}
        \frac{1}{2}[S^1,[S^1,\sum_MVMP_M+D\sum_MN_a^M]]+[S^2,\sum_MVMP_M+D\sum_MN_a^M]+[S^1,\sum_MT_{0,M}+\sum_{q\neq0,M}T_{q,M}]
\end{equation}
We choose $S^2$:
\begin{equation}
    [S^1,\sum_MT_{0,M}]+[S^2,\sum_MVMP_M+D\sum_MN_a^M]=0
\end{equation}
After projecting onto a fixed low-energy sector $P_M$, the second-order effective Hamiltonian is
\begin{equation}
V\sum_MP_MM+D\sum_MN_a^M+\sum_{M}T_{0,M}+\frac{1}{2}[S^1,\sum_{q\neq0,M}T_{q,M}]
\end{equation}

For the undoped insulator, $T_{0,0}=0$, and the nontrivial components of $S^1$ are
\begin{equation}
        S_{2,0}^1=\frac{T_{2,0}}{2V+D},S_{-2,2}^1=-\frac{T_{-2,2}}{2V+D}
\end{equation}
The second-order correction to the energy of the insulating state is therefore
\begin{equation}
        \delta E=-\frac{6Nt^2}{2V+D}
\end{equation}
where $N$ is the number of the enlarged unit cell.
\begin{figure}
    \centering
    \includegraphics[width=0.24\columnwidth]{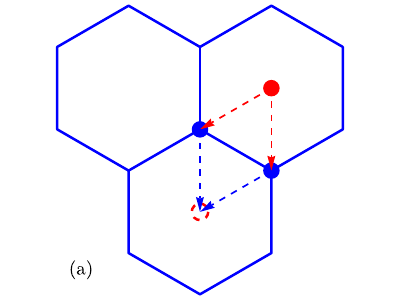}
    \includegraphics[width=0.24\columnwidth]{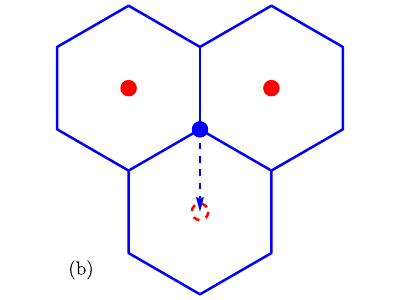}
     \caption{Effective hopping processes for doped charge carriers: (a) a particle hops between sites via two intermediate virtual states; (b) two neighboring particles coherently tunnel into a trimer configuration.}
    \label{strongcouplingdope}
\end{figure}

For a single doped electron, $M=z=6$, where $z$ is the coordination number. At the zeroth order, the single particle states are degenerate with energy $(6V+D)$. Virtual hopping processes away from the doped particle generate a second-order correction $\frac{-(6N-6)t^2}{2V+D}$. The remaining virtual processes generate an effective hopping of the doped particle, as shown in Fig.~\ref{strongcouplingdope}(a). The relevant matrix elements of the Schrieffer-Wolff generator are
\begin{equation}
        S^1_{-2,z+2}=\frac{T_{-2,z+2}}{D+6V-(2D+5V\times 2-3V)}=\frac{T_{-2,z+2}}{-V-D},S_{2,z}=\frac{T_{2,z}}{2D+5V\times2-3V-D-6V}=\frac{T_{2,z}}{V+D}
\end{equation}
Consequently, the second-order contribution within the one-particle sector is
\begin{equation}
        \frac{1}{2}S^1_{-2,z+2}T_{2,z}-\frac{1}{2}T_{-2,z+2}S^1_{2,z}
\end{equation}
which gives the effective nearest-neighbor hopping
\begin{equation}
-t_c\sum_{\langle ij\rangle }\left(c_i^\dagger c_j+c_j^\dagger c_i\right),t_c=\frac{t^2}{V+D}
\end{equation}
Thus, an isolated doped particle is mobile only through second-order virtual charge-transfer processes. 

For two doped electrons, $M=2z=12$. In this sector, the first-order hopping operator $T_{0,2z }$ is nontrivial when the two doped particles occupy neighboring plaquettes of the honeycomb lattice (see Fig.~\ref{strongcouplingdope}(b)). Since the plaquette centers form a triangular lattice, these configurations correspond to nearest-neighbor bond pairs on the triangular lattice. Such a bond pair can resonantly hybridize with a trimer configuration consisting of the two doped particles together with a particle-hole excitation of the insulating background. The Hilbert space spanned by the bond-pair and trimer configurations is closed under $T_{0,2z}$. In contrast, when the two doped particles are spatially separated, no hopping process acts within the degenerate manifold at first order. These states therefore retain the zeroth-order energy $2zV+2D$. The resonant bond-pair-trimer sector consequently provides a mechanism for lowering the energy of two doped particles already at first order in $t$. 

At $D=0$, the Hamiltonian within this sector has a chiral particle-hole symmetry: hopping connects only the bond-pair and trimer sectors, while the diagonal energies of the two sectors coincide. This produces a spectrum symmetric about zero and, in particular, a bound state below the energy of the separated-particle states. For general $D$, the spectrum can be solved in the momentum space. There are three inequivalent bond pairs and two trimers per triangular lattice unit cell.
\begin{equation}
        \begin{split}
        b_1^\dagger &= c_r^\dagger c_{r+a_1}^\dagger\\
        b_2^\dagger &= c_r^\dagger c_{r+a_2}^\dagger\\
        b_3^\dagger &= c_r^\dagger c_{r+a_2-a_1}^\dagger\\
        b_u^\dagger &=c_r^\dagger c_{r+a_1}^\dagger c_{r+a_2}^\dagger c_u\\
        b_d^\dagger &= c_r^\dagger c_{r+a_2}^\dagger c_{r+a_2-a_1}^\dagger c_d
        \end{split}
\end{equation}
In the basis $(b_1,b_2,b_3,b_u,b_d)$, the Hamiltonian at momentum $k$ is
\begin{equation}
       -t \begin{pmatrix}
                0&0&0&e^{-i5\pi/3}&-e^{-i2\pi/3}e^{ik(a_1-a_2)}\\
                0&0&0&-e^{-i3\pi/3}&e^{-i6\pi/3}\\
                0&0&0&e^{-i\pi/3}e^{-ika_1}&-e^{-i4\pi/3}\\
                e^{i5\pi/3}&-e^{i3\pi/3}&e^{i\pi/3}e^{ika_1}&-D/t&0\\
                -e^{i2\pi/3}e^{-ik(a_1-a_2)}&e^{i6\pi/3}&-e^{i4\pi/3}&0&-D/t
        \end{pmatrix}
\end{equation}
The energy and wavefunction of  ground-state are 
\begin{equation}
        E=\frac{1}{2}\left(D-\sqrt{D^2+24t^2}\right),|\psi\rangle = \sum_r\left(e^{i2\pi/3}b_1^\dagger +e^{i\pi/3}b_2^\dagger +b_3^\dagger \right)+\sum_r\frac{6te^{i\pi/3}}{D+\sqrt{D^2+24t^2}}\left(b_u^\dagger +b_d^\dagger \right)
\end{equation}
Taking the $\frac{t}{D}$ expansion, the energy $E=\frac{-6t^2}{D}$ and the wavefunction reduces to the two-body solution of the effective single-band model in the limit of $V\to\infty$. Thus, the chiral correlated-hopping mechanism of the effective model has a direct microscopic interpretation in terms of resonant hybridization between bond pairs and trimer configurations in the original three-band model.

\section{BCS mean-field approximation}
In this section, we investigate the superconducting ground state of the three-band model in the smaller charge-transfer-gap regime using the BCS mean-field approximation \cite{PhysRevB.105.094506}.

\begin{figure}
    \centering
    \includegraphics[width=0.24\columnwidth]{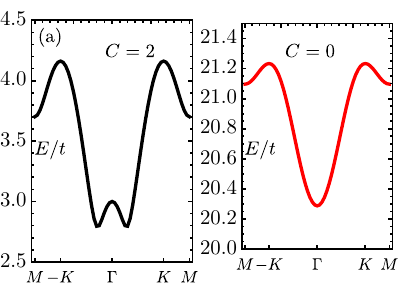}
    \includegraphics[width=0.24\columnwidth]{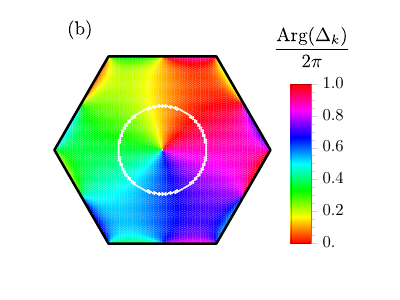}
    \includegraphics[width=0.24\columnwidth]{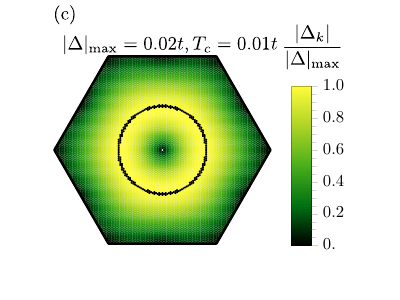}
     \caption{Mean-field solution for $D=2t,V=8t,x=0.2$ . (a) Quasiparticle dispersion of the top band without (left) and with (right) interaction corrections. (b) Phase and (c) magnitude of mean-field superconducting gap in momentum space.}
    \label{meanfield}
\end{figure}

Interaction effects can renormalize the normal-state band structure. In particular, the interaction-enhanced charge-transfer gap can render the top band topologically trivial. Within the Hartree–Fock approximation, the renormalized dispersion is given by
\begin{equation}
\begin{split}
    D&\to D+3V(n_{bc}-n_a)\\
    t&\to t+V\chi_1\\
    t'&\to t'+V\chi_2
    \end{split}
\end{equation}
where
\begin{equation}
    \begin{split}
        n_{bc}&=\frac{1}{2N}\sum_k(\rho_{bb}(k)+\rho_{cc}(k))\\
        n_a&=\frac{1}{N}\sum_k\rho_{aa}(k)\\
        \chi_1&=\frac{1}{6N}\text{Re}\sum_{k}\rho_{ca}(k)\times\left(e^{ik(a_1-a_2)}+e^{i2\pi/3}+e^{i4\pi/3}e^{-ika_2}\right)\\
        &+\frac{1}{6N}\text{Re}\sum_k\rho_{ba}(k)\times\left(e^{i\pi/3}+e^{i3\pi/3}e^{-ika_1}+e^{i5\pi/3}e^{-ika_2}\right)\\
        \chi_2&=\frac{1}{3N}\text{Re}\sum_k\rho_{bc}(k)\left(1+e^{ik(a_2-a_1)}+e^{-ika_1}\right)\\
        \rho_{ij}(k)&=\langle c_{k,j}^\dagger c_{k,i}\rangle
    \end{split}
\end{equation}
As shown in Fig.~\ref{meanfield}(a), interactions narrow the bandwidth and change the Chern number of the band intersected by the Fermi surface from $-2$ to $0$. 
The effective interactions between quasiparticles are obtained within the random-phase approximation (RPA). We introduce the vector of Fourier-transformed onsite density operators for the three orbitals,
\begin{equation}
    \begin{split}
    n_q&=(n_{q,a},n_{q,b},n_{q,c})\\
        n_{q,i}&=\sum_{k\in\text{BZ}}c_{k,i}^\dagger c_{k+q,i},i=a,b,c
    \end{split}
\end{equation}
The bare interaction can then be written in matrix form as 
\begin{equation}
    H_{\text{bare}}=\frac{1}{2N}\sum_{q\in\text{BZ}}n_q^TF(q)n_{-q}
\end{equation}
where 
\begin{equation}
    F(q)=V\begin{pmatrix}
        0&1+e^{iqa_1}+e^{iqa_2}&1+e^{iqa_2}+e^{-iq(a_1-a_2)}\\
        1+e^{-iqa_1}+e^{-iqa_2}&0&1+e^{-iqa_1}+e^{-iq(a_1-a_2)}\\
        1+e^{-iqa_2}+e^{iq(a_1-a_2)}&1+e^{iqa_1}+e^{iq(a_1-a_2)}&0
    \end{pmatrix}
\end{equation}
We define the static susceptibility matrix as
\begin{equation}
    \begin{split}
\chi_{ab}(q) &= \frac{1}{N\beta}\sum_{k\in\text{BZ},i\omega}G_{ab}(k,i\omega)G_{ba}(k+q,i\omega)\\
&=\frac{1}{N}\sum_{k\in\text{BZ}}\sum_{n,m}\frac{n_f(E_n(k))-n_f(E_m(k+q))}{E_n(k)-E_m(k+q)}u_{an}(k)\bar{u}_{bn}(k)u_{bm}(k+q)\bar{u}_{am}(k+q)
    \end{split}
\end{equation}
where $G_{ab}(k,i\omega)$ is the single-particle Green's function, $E_n(k)$ and $u_{an}(k)$ are the energy and orbital-$a$ component of band-$n$ Bloch state, respectively, and $n_f(E)$ is Fermi-Dirac distribution. 
The susceptibilities $\bar{\chi}_{ab}(q)=\chi_{ba}(q)=\chi_{ab}(-q)$.
Within the RPA , the effective interaction is
\begin{equation}
    \begin{split}
    H_{\text{RPA}}&=\frac{1}{2N}\sum_{q\in\text{BZ}}n_{q}^TF(q)\frac{1}{I-\chi(q)F(q)}n_{-q}\\
    &\equiv\frac{1}{2N}\sum_{q\in\text{BZ}}n_q^TV_{\text{eff}}(q)n_{-q}
    \end{split}
\end{equation}
where $I$ is the identity matrix.  
We project the RPA-renormalized interaction onto the BCS pairing channel, assuming that pairing occurs only within the $n$-th band, which crosses the Fermi surface. The resulting interaction is
\begin{equation}
    \begin{split}
H_{\text{BCS}}&=\frac{1}{2N}\sum_{k_1,k_2\in\text{BZ}}c_{k_1,n}^\dagger c_{-k_1,n}^\dagger c_{-k_2,n}c_{k_2,n}\bar{u}_{an}(k_1)\bar{u}_{bn}(-k_1)u_{bn}(-k_2)u_{an}(k_2)V(k_2-k_1)_{ab}\\
&=\frac{1}{2N}\sum_{k_1,k_2\in\text{BZ}}c_{k_1}^\dagger c_{-k_1}^\dagger c_{-k_2}c_{k_2}\Gamma(k_2,k_1)\\
\Gamma(k_2,k_1)&\equiv\bar{u}_{an}(k_1)\bar{u}_{bn}(-k_1)u_{bn}(-k_2)u_{an}(k_2)V_{\text{eff}}(k_2-k_1)_{ab}\\
\Gamma(k_2,-k_1)&\equiv\bar{u}_{an}(-k_1)\bar{u}_{bn}(k_1)u_{bn}(-k_2)u_{an}(k_2)V_{\text{eff}}(k_2+k_1)_{ab}
    \end{split}
\end{equation}
where $c_{k,n}\equiv c_k$ annihilates a fermion in the $n$-th band and $\Gamma(k_2,k_1)$ is the projected pairing vertex.
The Bloch wavefunction has a $U(1)$ gauge redundancy under
\begin{equation}
    u_n(k)\to u_n(k)\times e^{i\theta(k)},c_k\to c_k e^{-i\theta(k)}
\end{equation}
For a topologically trivial band localized on $1a$ sites, we fix this gauge by requiring the $1a$-component of $u_{n}(k)$ to be real and positive throughout the Brillouin zone.
 Antisymmetrizing the pairing interaction and restricting the momentum sums to the half Brillouin zone $\frac{1}{2}\text{BZ}$, we obtain
\begin{equation}
    H_{\text{BCS}}=\frac{1}{N}\sum_{k_1,k_2\in\frac{1}{2}\text{BZ}}c_{k_1}^\dagger c_{-k_1}^\dagger c_{-k_2}c_{k_2}\Gamma^A(k_2,k_1)
\end{equation}
where the antisymmetrized pairing vertex is
\begin{equation}
    \Gamma^A(k_2,k_1)= \Gamma(k_2,k_1)-\Gamma(k_2,-k_1)
\end{equation}
The resulting kernel $\Gamma^A$ is Hermitian in the space of Cooper pair momenta. 
At zero temperature, the mean-field decoupling of the BCS interaction yields 
\begin{equation}
    H_{\text{MF}}=\sum_{k_1\in\frac{1}{2}\text{BZ}}c_{k_1}^\dagger c_{-k_1}^\dagger \Delta(k_1)+\text{H.c.}
\end{equation}
where the gap function is determined self-consistently by
\begin{equation}
    \Delta(k_1)=\frac{1}{N}\sum_{k_2\in\frac{1}{2}\text{BZ}}\langle c_{-k_2}c_{k_2}\rangle \Gamma^A(k_2,k_1)
\end{equation}
The BdG Hamiltonian is 
\begin{equation}
    H_{\text{BdG}}=\sum_{k\in\frac{1}{2}\text{BZ}}
    \begin{pmatrix}
        c_{k}^\dagger & c_{-k}
    \end{pmatrix}
    \begin{pmatrix}
        \epsilon_k-\mu&\Delta(k)\\
        \Delta(k)^*&\mu-\epsilon_k
    \end{pmatrix}
    \begin{pmatrix}
        c_k\\
        c_{-k}^\dagger
    \end{pmatrix}
\end{equation}
The Hamiltonian is diagonalized by the unitary transformation
\begin{equation}
    \begin{pmatrix}
        c_k\\
        c_{-k}^\dagger 
    \end{pmatrix}
    =U_k\begin{pmatrix}
        \gamma_k\\
        \gamma_{-k}^\dagger 
    \end{pmatrix}
\end{equation}
with 
\begin{equation}
    \begin{split}
    U_k&=\begin{pmatrix}
        u_k&-\bar{v}_k\\
        v_k&\bar{u}_k
    \end{pmatrix},E_k=\sqrt{\xi_k^2+|\Delta_k|^2},\xi_k=\epsilon_k-\mu\\
    u_k&=\frac{E_k+\xi_k}{\sqrt{2E_k(E_k+\xi_k)}},v_k=\frac{\bar{\Delta}_k}{\sqrt{2E_k(E_k+\xi_k)}}
\end{split}
\end{equation}
At zero temperature, the anomalous expectation value is
\begin{equation}
    \langle c_{-k}c_k\rangle =\bar{v}_ku_k\langle \gamma_k^\dagger \gamma_k\rangle -u_k\bar{v}_k\langle \gamma_{-k}\gamma_{-k}^\dagger \rangle=-u_k\bar{v}_k=\frac{-\Delta_k}{2E_k}
\end{equation}
The momentum space occupation is 
\begin{equation}
    \langle c_k^\dagger c_k\rangle=\langle c_{-k}^\dagger c_{-k}\rangle = |v_k|^2=\frac{1}{2}\left(1-\frac{\xi_k}{E_k}\right)
\end{equation}
Finally, the chemical potential is determined by the filling constraint
\begin{equation}
    \frac{1}{N}\sum_{k\in\frac{1}{2}\text{BZ}}\left(1-\frac{\xi_k}{E_k}\right)=x
\end{equation}
The zero-temperature gap equation is solved self-consistently, and the resulting gap function is shown in Fig.~\ref{meanfield}(b)(c).
Its phase exhibits a characteristic $(p+ip)$-wave winding around the $\Gamma$ point. The Fermi surface consists of a single pocket centered at $\Gamma$, where the gap vanishes, while the gap magnitude is maximal along the Fermi surface. Consequently, the quasiparticle spectrum is fully gapped. The resulting superconducting state has a BdG Chern number $C=1$. 

\section{Chern numbers of Bloch bands and BdG Hamiltonian}
Given the Bloch states $u(k)$ on a momentum-space grid, we define the Berry connection $A(k)$ along the $k_2$ direction through the overlap between neighboring states:
\begin{equation}
    \langle u(k_1,k_2+dK_2)|u(k_1,k_2)\rangle \approx 1-\langle u(k)|\partial_2|u(k)\rangle dK_2\approx e^{-i A_2(k)dK_2}
\end{equation}
Under the gauge redundancy $|u(k)\rangle \to |u(k)\rangle e^{i\theta(k)}$, the overlap between neighboring Bloch states transform as 
\begin{equation}
    \langle u(k_1,k_2+dK_2)|u(k_1,k_2)\rangle\to \langle u(k_1,k_2+dK_2)|u(k_1,k_2)\rangle\times e^{i\theta(k_1,k_2)-i\theta(k_1,k_2+dK_2)}
\end{equation}
The Wilson loop along the $k_2$ direction at fixed $k_1$ is defined as the product of normalized overlaps between neighboring Bloch states around a closed loop in the Brillouin zone
\begin{equation}
    W_2(k_1)=\prod_{k_2}\frac{\langle u(k_1,k_2+dK_2)|u(k_1,k_2)\rangle}{|\langle u(k_1,k_2+dK_2)|u(k_1,k_2)\rangle|}\approx \exp\left(-i\int_{k_1\vec{b}_1}^{k_1\vec{b}_1+\vec{b}_2}\vec{A}(k)d\vec{k}\right)
\end{equation}
The Chern number can be obtained from the winding of the Wilson loop as $k_1$ traverses the Brillouin zone.
\begin{equation}
    \begin{split}
    -id\ln W_2(k_1)&=\int_{k_1\vec{b}_1}^{k_1\vec{b}_1+\vec{b}_2}\vec{A}(k)d\vec{k}-\int_{(k_1+dk_1)\vec{b}_1}^{(k_1+dk_1)\vec{b}_1+\vec{b}_2}\vec{A}(k)d\vec{k}\\
    -i\int_0^1d\ln W_2(k_1)&=-\int_{\text{BZ}}\Omega(\vec{k})d^2k=-2\pi C
    \end{split}
\end{equation}
where $\Omega(k)$ is the Berry curvature, whose integration over the Brillouin zone gives the Chern number $C$ with factor $2\pi$.
The BdG Hamiltonian $H_{\text{BdG}}(k)$ can be extended from the half Brillouin zone to the full Brillouin zone  by defining
\begin{equation}
H_{\text{BdG}}(-k)=    \begin{pmatrix}
        \xi_{-k}&\Delta(-k)\\
        \bar{\Delta}(-k)&-\xi_k
        \end{pmatrix}
\end{equation}
where Fermion statistics require the gap function to satisfy $\Delta(-k)\equiv-\Delta(k)$ . The BdG Chern number is defined as the Chern number of the lower-energy Bogoliubov band over the full Brillouin zone.
\begin{figure}
    \centering
    \includegraphics[width=0.24\columnwidth]{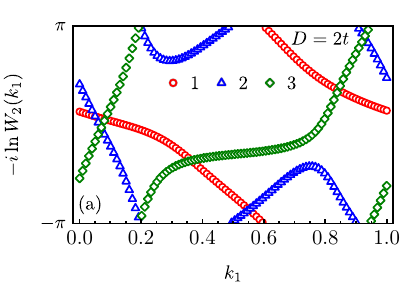}
    \includegraphics[width=0.24\columnwidth]{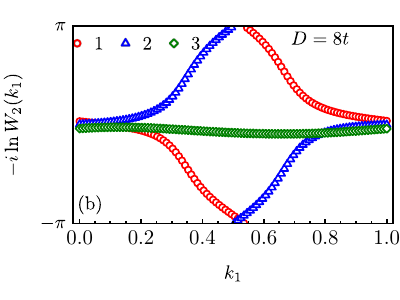}
    \includegraphics[width=0.24\columnwidth]{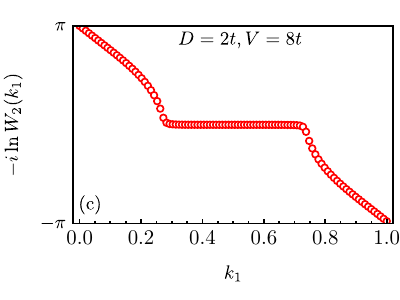}
     \caption{Wilson-loop spectra. The band energies are ordered as $E_1<E_2<E_3$ (a) Noninteracting bands in the smaller-$D$ regime. (b) Noninteracting bands in the large-$D$ regim. (c) BdG Hamiltonian.}
    \label{wilson}
\end{figure}
Figure.~\ref{wilson} shows the Wilson-loop spectra of the noninteracting Bloch Hamiltonian and the BdG Hamiltonian, from which we determine the Chern numbers reported in the main text.

\section{Hall metal}
\begin{figure}
    \centering
    \includegraphics[width=0.24\columnwidth]{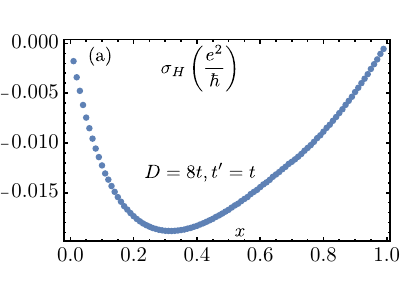}
    \includegraphics[width=0.24\columnwidth]{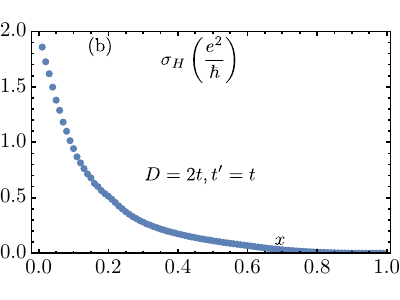}
    \includegraphics[width=0.24\columnwidth]{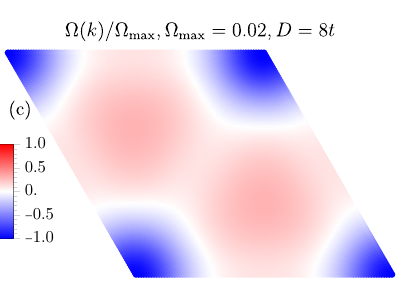}
    \includegraphics[width=0.24\columnwidth]{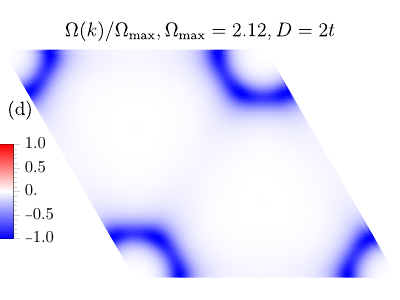}
     \caption{Doping dependence of Hall conductance for (a) $D=8t$ and (b) $D=2t$, together with the corresponding Berry-curvature distributions of the top band in (c) and (d), respectively.}
    \label{hallandberry}
\end{figure}
For noninteracting fermions at a generic filling, the Hall conductance is given by
\begin{equation}
    \sigma_{H}=\frac{e^2}{\hbar}\frac{1}{A}\sum_{n,\vec{k}\in\text{BZ}}\Omega_n(k)f(\beta(E_n(\vec{k}-\mu))),f(x)=\frac{1}{1+e^x}
\end{equation}
where $A$ is the system area, $\Omega_n(\vec{k})$ is the Berry curvature of the $n$-th band, and $f(x)$ is the Fermi-Dirac distribution.
For the charge-transfer insulator at $D=8t$, the doping dependence of $\sigma_H$ is shown in Fig.~\ref{hallandberry}(a), and the Berry-curvature distribution of the top band is shown in Fig.~\ref{hallandberry}(c). The undoped insulator at $x=0$ is topologically trivial and therefore has vanishing $\sigma_H$. Upon doing, the system becomes metallic and exhibits a negative Hall conductance. For comparison, Fig.~\ref{hallandberry}(b) and (d) show the Hall conductance and Berry-curvature distribution, respectively, for $D=2t$, where the doped band has Chern number $C=-2$. In this case, the Berry curvature exhibits a ring-shaped distribution centered at the $\Gamma$ point, and the undoped insulator has a quantized Hall conductance of magnitude $2\frac{e^2}{\hbar}$.

\section{Correlated hopping as a pairing mechanism}
\begin{figure}
    \centering
    \includegraphics[width=0.24\columnwidth]{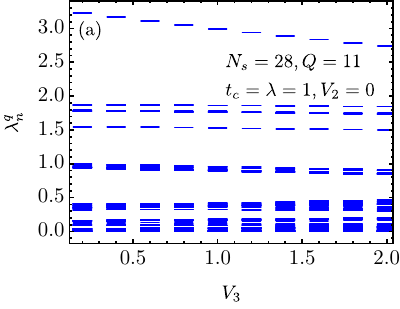}
    \includegraphics[width=0.24\columnwidth]{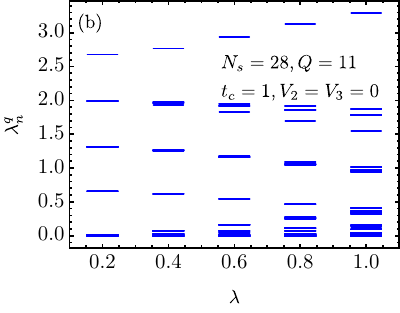}
    \includegraphics[width=0.24\columnwidth]{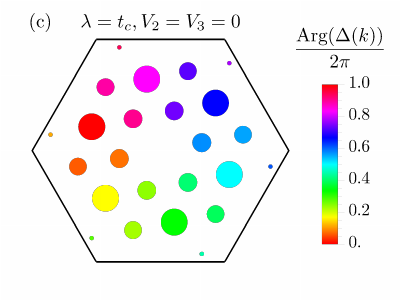}
    \includegraphics[width=0.24\columnwidth]{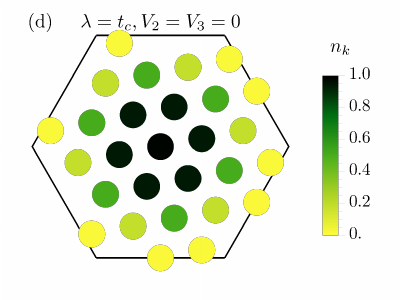}
     \caption{Two-particle density-matrix spectra as functions of $V_3$ (a) and $\lambda$ (b). (c) Pairing wavefunction and (d) momentum space occupation distribution for the effective Hamiltonian with correlated hopping as sole interaction term. All calculations are done on a torus with $28$ sites and $11$ particles.}
    \label{correlatedhopping}
\end{figure}
The effective single-band Hamiltonian is 
\begin{equation}
        H_{\text{eff}}=-t_c\sum_{\langle ij\rangle }\left(c_i^\dagger c_j+h.c.\right)
         + V_2\sum_{\langle ij\rangle}n_in_j+V_3\sum_{ijk\in\triangle/\triangledown}n_in_jn_k+\lambda\sum_{ijk\in\triangle/\triangledown}P_{\text{cyc}}\left( n_ic_k^\dagger c_je^{i\phi_{kj}} + h.c. \right)
\end{equation}
where the effective couplings depend on the microscopic interaction parameters. For $V=D$, both the chiral correlated hopping amplitude $\lambda$ and the three-body interaction $V_3$ are nonzero. To investigate the role of correlated hopping in pairing, we independently tune these effective couplings.

We first fix $t_c=\lambda=1,V_2=0$, and calculate the two-particle density-matrix spectrum as a function of $V_3$. As shown in Fig.~\ref{correlatedhopping}(a), the leading eigenvalue increases as $V_3$ decreases, indicating an enhancement of pairing correlations as the three-body interaction is reduced. Conversely, reducing $\lambda$ decreases the leading eigenvalue [Fig.~\ref{correlatedhopping}(b)], highlighting the important role of correlated-hopping in prompting pairing. 

We further find that the leading eigenvector belongs to the Cooper-pair momentum sector $q=(0,0)$. Its momentum space structure exhibits the characteristic $(p+ip)$-wave phase winding structure [Fig.~\ref{correlatedhopping}(c)], while the occupation number distribution in momentum space shows a single Fermi pocket centered at the $\Gamma$ point [Fig.~\ref{correlatedhopping}(d)]. Together, these results support chiral correlated hopping as a pairing mechanism in the effective single-band model. The resulting superconducting state has a BdG Chern number $C=1$.

\section{Other DMRG results}
\subsection{Convergence of DMRG calculation}
\begin{figure}
    \centering
    \includegraphics[width=0.24\columnwidth]{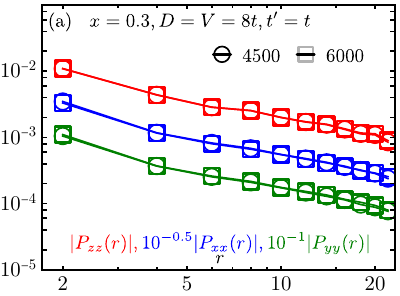}
    \includegraphics[width=0.24\columnwidth]{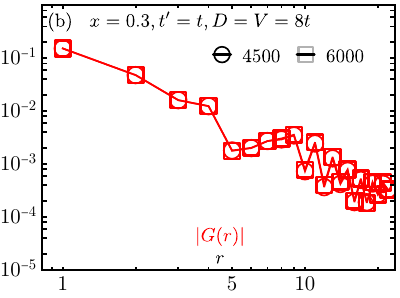}
    \includegraphics[width=0.24\columnwidth]{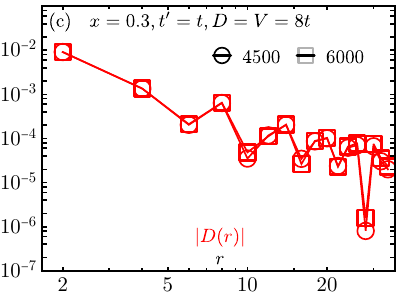}
     \caption{Comparing results obtained with two different maximum bond dimensions.}
    \label{convergence}
\end{figure}

The matrix-product state in the DMRG calculation is limited by the maximum bond dimension. In Fig.~\ref{convergence}, we demonstrate the convergence of the correlation functions by comparing results obtained with two different maximum bond dimensions.

\subsection{Superconducting pairing correlations of different components}
\begin{figure}
    \centering
        \includegraphics[width=0.24\columnwidth]{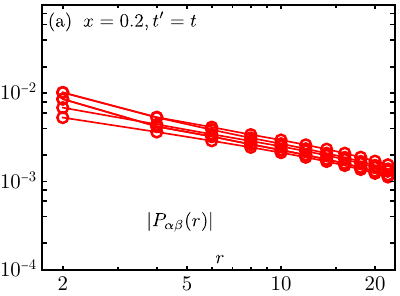}
        \hfill
        \includegraphics[width=0.24\columnwidth]{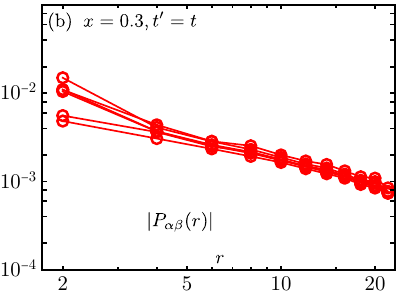}
        \hfill
        \includegraphics[width=0.24\columnwidth]{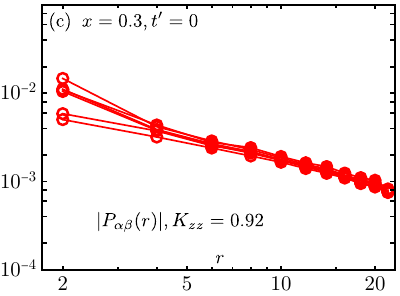}
        \hfill
        \includegraphics[width=0.24\columnwidth]{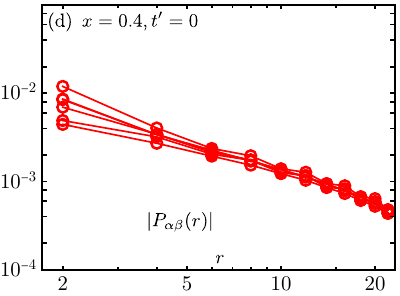}        
        \caption{The pairing correlations for the model with $D=V=8t$.}
        \label{largeDallsc}
\end{figure}

\begin{figure}
    \centering
    \includegraphics[width=0.24\columnwidth]{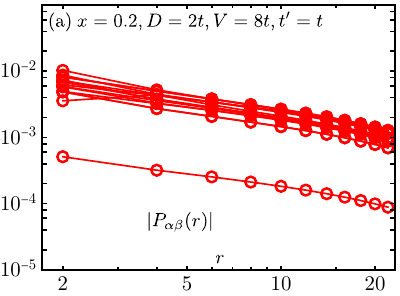}
    \includegraphics[width=0.24\columnwidth]{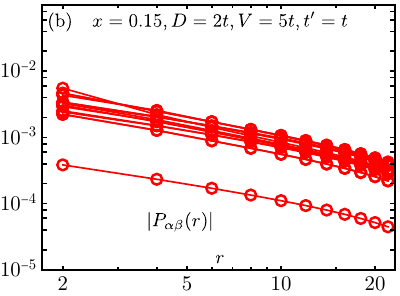}
    \includegraphics[width=0.24\columnwidth]{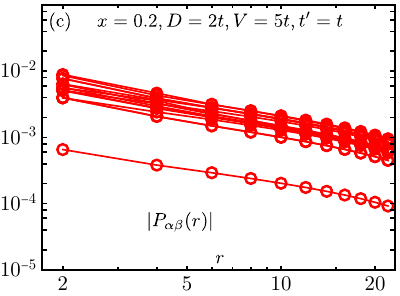}
     \caption{Pairing correlations in smaller-gap regime where $D=2t$.}
    \label{smallDallsc}
\end{figure}

As shown in Fig.~\ref{largeDallsc}, pairing correlations develop between different near-neighbor bonds in the large charge-transfer-gap regime. The comparable magnitudes of the correlations for different bond orientations indicate that the superconducting correlations are nearly isotropic.
In Fig.~\ref{smallDallsc}, we include the pairing correlations from trimer operators in the samller charge-transfer-gap regime in addition to those from bond pairs. 

\subsection{The edge structure in single-particle sector}
\begin{figure}
    \centering
 \includegraphics[width=0.24\columnwidth]{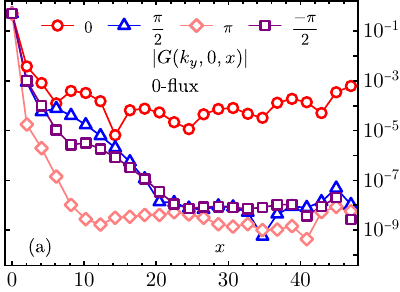}
        \includegraphics[width=0.24\columnwidth]{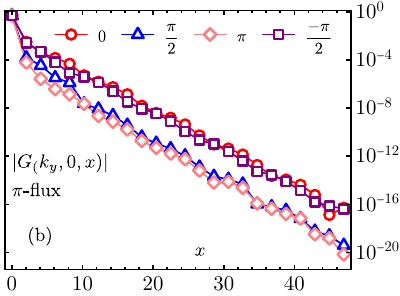}
        \caption{Momentum-resolved single-particle correlations with one point fixed at the left edge. (a) and (b) correspond to different flux-sectors. The different symbols denote different $k_y$ channels.}
        \label{edge}
\end{figure}

In the Kitaev chain, the nontrivial topology phase is characterized by a Majorana zero mode localized at each open edge. This naturally raises the question of how the corresponding topological edge structure manifests itself in a charge-conserving system with superconducting phase fluctuations. At the mean-field level, the Bogoliubov--de Gennes (BdG) Hamiltonian describes the single-particle sector, suggesting that the single-particle Green's function provides a natural many-body diagnostic. We resolve the Green's function into momentum channels along the periodic $y$-direction by Fourier transformation,
\begin{equation}
        G(k_y,r_0,x)=\sum_{j=0}^{L_y-1}\frac{1}{\sqrt{L_y}}\exp(i k_y j)G(r_0,(x,j))
\end{equation}
where $j$ denotes the coordinate along the $y$-direction and $k_y\in\frac{2\pi}{L_y}\mathbb{Z}$. We fix $r_0$ at the left edge of the cylinder and examine the single-particle correlations between this edge point and sites at different positions $x$. As shown in Fig.~\ref{edge}(a), most momentum channels decay rapidly as the separation from the left edge increases. In the Luttinger liquid regime, however, the $k_y=0$ component exhibits a distinct nonlocal behavior: although it initially decays away from the left edge, it increases again upon approaching the opposite edge. By contrast, in the strongly paired phase [Fig.~\ref{edge} (b)], all momentum channels decay into the bulk without any appreciable enhancement near the remote edge. This contrast indicates that the Luttinger-liquid phase supports a nonlocal single-particle correlation connecting the two edges, consistent with the edge structure associated with the underlying topological phase.

\subsection{Numerical results of $L_y=3$ geometry}
In this subsection, we present DMRG results for systems with $L_y=3$ in both the large- and small-charge-transfer-gap regimes. 
\begin{figure}
    \centering
 \includegraphics[width=0.24\columnwidth]{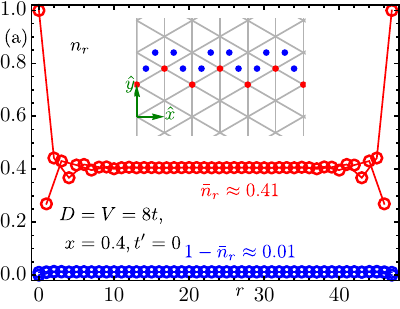}
 \includegraphics[width=0.24\columnwidth]{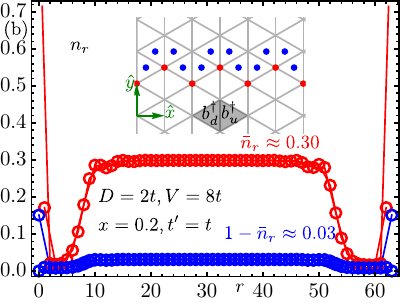}
\includegraphics[width=0.24\columnwidth]{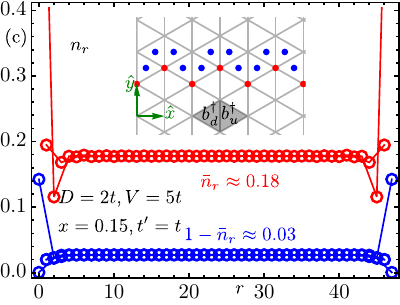}
\includegraphics[width=0.24\columnwidth]{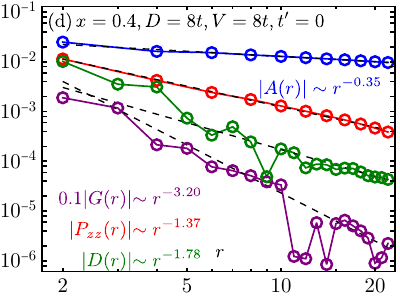}
\vspace{2mm}
\includegraphics[width=0.24\columnwidth]{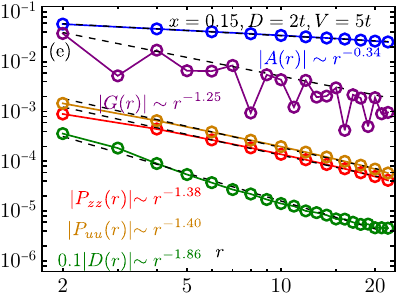}
\includegraphics[width=0.24\columnwidth]{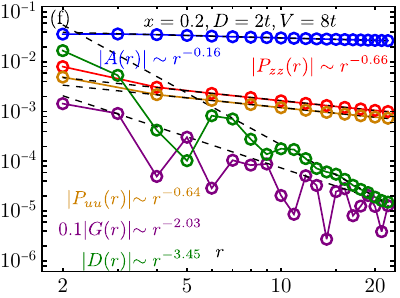}
\includegraphics[width=0.24\columnwidth]{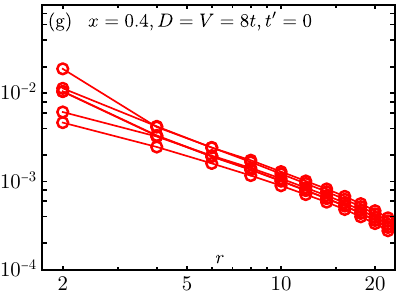}
\includegraphics[width=0.24\columnwidth]{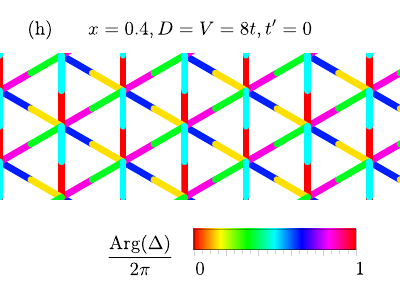}
\caption{DMRG results for systems with $L_y=3$.}
        \label{Ly3}
\end{figure}
As shown in Fig.~\ref{Ly3}(a-c), the charge density remains uniform within each sublattice in both regimes, with no evidence of static charge order.
The measured correlation functions exhibit power-law decay [Fig.~\ref{Ly3}(d-f)]. The scaling exponents of the nonlocal string correlations ($K_{s}$) and superconducting pairing correlations ($K_{\text{SC}}$) satisfy $K_{s}=\frac{1}{4}K_{\text{SC}}$, implying a nontrivial topology in the single-particle sector. The systems have quasi-long-ranged pairing order with exponents $K_{\text{SC}}<2$. Moreover, the pairing correlation amplitudes are comparable across different bond orientations [Fig.~\ref{Ly3}(g)], indicating nearly isotropic pairing correlations, while their relative phases follow the characteristic $p+ip$-wave pattern [Fig.~\ref{Ly3}(h)].

\section{Technical details of exact-diagonalization}

\subsection{Torus geometry}
We perform exact diagonalization of the effective single-band model on a triangular lattice torus. The lattice sites are labeled by integer-coordinated vectors based on unit vectors $\vec{a}_1$ and $\vec{a}_2$: 
\begin{equation}
        \vec{r}=r_1\vec{a}_1+r_2\vec{a}_2,r_1,r_2\in\mathbb{Z}
\end{equation}
A finite torus is defined by identifying lattice points related by two linearly independent integer translation vectors,
\begin{equation}
        r=(r_1,r_2),r\sim r+(n_{11},n_{12}),r\sim r+(n_{21},n_{22}),
        n=\begin{pmatrix}
                n_{11}&n_{12}\\
                n_{21}&n_{22}
        \end{pmatrix}
\end{equation}
By applying a unimodular integer transformation $U$, the matrix $n$ can be brought to Hermite normal form, 
\begin{equation}
        Un = \begin{pmatrix}
                L_1&m\\
                0&L_2
        \end{pmatrix}
\end{equation}
where $L_1,L_2>0$ and $0\leq m < L_1$. Thus, without loss of generality, the torus can be specified by the periodic identifications $r\sim r+(L_1,m),r\sim r+(0, L_2)$. A comprehensive classification of finite-size torus geometries and their spatial symmetries can be found in \cite{wilhelm2021Interplay}.
The allowed crystal momenta are determined by requiring the plane-wave factor $e^{i\vec{k}\cdot\vec{r}}$ to be invariant under the two periodic translations. The resulting momenta form the reciprocal lattice generated by
\begin{equation}
        \vec{k}=\begin{pmatrix}
                k_1&k_2
        \end{pmatrix}
        \begin{pmatrix}
                \frac{1}{L_1}&0\\
                \frac{-m}{L_1L_2}&\frac{1}{L_2}
        \end{pmatrix}
        \begin{pmatrix}
                \vec{b}_1\\
                \vec{b}_2
        \end{pmatrix},k_1,k_2\in \mathbb{Z}
\end{equation}
where $\vec{b}_i$ are the reciprocal lattice vectors satisfying $\vec{a}_i\cdot\vec{b}_j=2\pi\delta_{ij}$. In our calculations, we employ the following torus geometries:
\begin{equation}
    (L_1,L_2,m)=(1,21,4),(5,5,0),(3,9,3),(2,14,4),(7,7,0)
\end{equation}
They are derived from following integer matrices, respectively
\begin{equation}
    n=\begin{pmatrix}
        1&4\\
        5&-1
    \end{pmatrix},
    \begin{pmatrix}
        5&0\\
        0&5
    \end{pmatrix},
    \begin{pmatrix}
        3&3\\
        3&-6
    \end{pmatrix},
    \begin{pmatrix}
        2&4\\
        6&-2
    \end{pmatrix},
    \begin{pmatrix}
        7&0\\
        0&7
    \end{pmatrix}
\end{equation}

\subsection{Many-body Hilbert space}
The model conserves the total charge and is invariant under lattice translations. We construct the many-body Hilbert space from real-space Fock states with fixed total charge. These states decompose into disconnected translation orbits under the action of the translation group. Within each orbit, all states are related by lattice translations. Using the binary representation of the Fock states, we uniquely choose a representative $|\sigma\rangle$ for each orbit as the state whose binary number is the smallest within that orbit. The translation eigenstates can then be constructed as 
\begin{equation}
        \begin{split}
        |k,\sigma\rangle &=\sum_r T^r|\sigma\rangle\chi_k(r),k=(k_1,k_2),r=(r_1,r_2),T^r=T_1^{r_1}T_2^{r_2}\\
        \chi_k(r)&=\exp\left(i2\pi
        \begin{pmatrix}
                k_1&k_2
        \end{pmatrix}
        \begin{pmatrix}
                \frac{1}{L_1}&0\\
                \frac{-m}{L_1L_2}&\frac{1}{L_2}
        \end{pmatrix}
        \begin{pmatrix}
                r_1\\
                r_2
        \end{pmatrix}
        \right)\\
        T^r|k,\sigma\rangle& = \chi_k(-r)|k,\sigma\rangle
        \end{split}
\end{equation}
All nonvanishing states obtained in this way form a complete and linearly independent basis of the many-body Hilbert space at fixed charge and crystal momentum.
The norm of a constructed state is 
\begin{equation}
        \langle k,\sigma|k,\sigma\rangle =\sum_r\langle \sigma| T^{-r}|k,\sigma\rangle \chi_k(-r)=N\langle \sigma|k,\sigma\rangle=N\sum_{r\in S_\sigma}\eta_{\sigma,r}\chi_k(r) 
\end{equation}
where $S_{\sigma}$ denotes the stabilizer subgroup of $|\sigma\rangle$, i.e., the set of translations that leave the Fock configuration invariant up to the fermionic exchange sign. Here, $\eta_{\sigma,r}=\pm1$ is the fermion sign generated by the action of $T^r$ on $|\sigma\rangle$. For $r$ in $S_\sigma$, the action of a stabilizer translation is 
\begin{equation}
        T^r|k,\sigma\rangle = \eta_{\sigma,r}|k,\sigma\rangle = \chi_k(-r)|k,\sigma\rangle,r\in S_{\sigma}
\end{equation}
Therefore, a sufficient and necessary condition for $|k,\sigma\rangle$ to be nonvanishing is $\eta_{\sigma,r}\chi_k(r)=1,\forall r\in S_\sigma$. The norm of $|k,\sigma\rangle$ is $\langle k,\sigma|k,\sigma\rangle= N|S_\sigma|$. A useful identity is
\begin{equation}
        \langle \sigma|k,\sigma\rangle = \sum_{r\in S_\sigma}\eta_{\sigma,r}\chi_k(r)=|S_\sigma|
\end{equation}

\subsection{Representation of Hamiltonian}
The Hamiltonian can be decomposed into a sum of local operators related by lattice translations. It is therefore sufficient to consider the action of each local operator on a representative Fock state. By construction, the action of a local operator $H_i$ on $|\sigma\rangle$ generates a single Fock state, which can be related to another representative state $|\tau_{i,\sigma}\rangle$ by a lattice translation. We write
\begin{equation}
        \begin{split}
        H|k,\sigma\rangle_n &= \frac{1}{\sqrt{N|S_\sigma|}}\sum_rT^rH|\sigma\rangle \chi_k(r)\\
                &=\frac{1}{\sqrt{N|S_\sigma|}}\sum_rT^r\sum_iH_i|\sigma\rangle \chi_k(r)\\
                &=\frac{1}{\sqrt{N|S_\sigma|}}\sum_rT^r\sum_ih_iT^{-d_{i,\sigma}}\eta_{\tau,d}|\tau_{i,\sigma}\rangle \chi_k(r)\\
                &=\sum_ih_i\eta_{\tau,d}|k,\tau_{i,\sigma}\rangle_n \chi_k(d_{i,\sigma})\sqrt{\frac{|S_\tau|}{|S_\sigma|}}
        \end{split}
\end{equation}
where $|k,\sigma\rangle_n$ denotes the normalized translation eigenstate $|k,\sigma\rangle$, $h_i$ is the matrix elements of $H_i$ between Fock states, and $d$ is the translation connecting to the representative state $|\tau\rangle$, whose action sign is $\eta$.

\subsection{Flux insertion}
Flux can be inserted through the torus by introducing a uniform gauge potential $\vec{A}=A_1\vec{b}_1+A_2\vec{b}_2$. The gauge-invariant fluxes threading the two noncontractible cycles of the torus are
\begin{equation}
        \begin{pmatrix}
                \theta_1\\
                \theta_2
        \end{pmatrix}=2\pi \begin{pmatrix}
                n_{11}&n_{12}\\
                n_{21}&n_{22}
        \end{pmatrix}\begin{pmatrix}
                A_1\\
                A_2
        \end{pmatrix}
\end{equation}
Consequently, a hopping process along $\vec{r}=r_1\vec{a}_1+r_2\vec{a}_2$ acquires an additional Peierls phase 
\begin{equation}
\exp(i\vec{A}\cdot \vec{r})=\exp\left(i
        \begin{pmatrix}
             r_1&r_2   
        \end{pmatrix}
        n^{-1}
        \begin{pmatrix}
                \theta_1\\
                \theta_2
        \end{pmatrix}
\right)
\end{equation}

\subsection{Observables measurement}
On a finite system, the ground state can be chosen as an eigenstate of the translation operators. Consequently, the expectation value of an operator with nonzero crystal momentum vanishes, and it is sufficient to consider translation-invariant operators. Such operators can be written in the form
\begin{equation}
        \sum_R\langle O_R\rangle
\end{equation}
where $O_R$ denotes the translation of a local operator to position $R$.
We describe how the expectation value of $O$ can be evaluated directly in the symmetry-reduced basis.
Without loss of generality, suppose that the action of $O_R$ on a representative product state $|\sigma\rangle$ produces a single product state,
\begin{equation}
        \begin{split}
     \sum_R\langle O_R\rangle &=\sum_{R,s,\sigma}\langle G|T^sO_R|\sigma\rangle \frac{\chi_k(s)}{\sqrt{N|S_\sigma|}}\psi_G(\sigma)\\
     &=\sum_{R,\sigma}\langle G|O_R|\sigma\rangle \sqrt{\frac{N}{|S_\sigma|}}\psi_G(\sigma)\\
     &=\sum_{R,\sigma}\langle G|\eta_{R,\sigma}T^{-d_{R,\sigma}}|\tau_{R,\sigma}\rangle \sqrt{\frac{N}{|S_\sigma|}}\psi_G(\sigma)\\
     &=\sum_{R,\sigma}\eta_{R,\sigma}\langle G|\tau_{R,\sigma}\rangle \chi_k(d_{R,\sigma})\sqrt{\frac{N}{|S_\sigma|}}\psi_G(\sigma)\\
     &=\sum_{R,\sigma}\eta_{R,\sigma}\bar{\psi}_G(\tau)\langle \tau,k|\tau\rangle\chi_k(d_{R,\sigma})\sqrt{\frac{1}{|S_\tau||S_\sigma|}}\psi_G(\sigma)\\
     &=\sum_{R,\sigma}\bar{\psi}_G(\tau)\eta_{R,\sigma}\sqrt{\frac{|S_\tau|}{|S_\sigma|}}\chi_k(d_{R,\sigma})\psi_G(\sigma)
        \end{split}  
\end{equation}
where $\eta_{R,\sigma}$ is the corresponding matrix element, $d_{R,\sigma}$ is the translation required to bring the resulting state back to its representative state, and $|\tau_{R,\sigma}\rangle$ is the resulting representative product state. As an example, the single-particle occupation in momentum space is 
\begin{equation}
        n_k=\langle c_k^\dagger c_k\rangle =\frac{1}{N}\sum_{R,r}\langle c_{R+r}^\dagger c_R\rangle e^{ikr}
\end{equation}
We can first evaluate the translation-summed real-space Green function and subsequently obtain the momentum distribution through the Fourier transformation.
\begin{equation}
G(r)=\sum_R\langle c_{R+r}^\dagger c_R\rangle
\end{equation}
Similarly, the two-particle density matrix at center-of-mass momentum $q$ is defined as
\begin{equation}
        \begin{split}
        \rho^q_{k_1,k_2}&=\langle c_{k_1+q}^\dagger c_{-k_1}^\dagger c_{-k_2}c_{k_2+q}\rangle\\
        &=\frac{1}{N^2}\sum_{r_1,r_2,r_3}\sum_R\langle c_{R+r_3}^\dagger c_{R+r_2}^\dagger c_{R+r_1}c_R\rangle e^{ik_2r_1-ik_1r_2+i(k_1+q)r_3}
        \end{split}
\end{equation}
Here we can take $O_R$ to be $c_{R+r_3}^\dagger c_{R+r_2}^\dagger c_{R+r_1}c_R$.

\section{String operator for the one-dimensional spinless fermion model}
In this section, we study the phase-fluctuating one-dimensional $p$-wave superconducting model introduced in \cite{Kane2017}. This model has two phases which at the mean-field level are equivalent to two phases of the Kitaev chain with different topology. Using a duality transformation, one can establish a relation between the scaling dimensions of the charge-$1e$ operators connected by the Jordan-Wigner string and those of the charge-$2e$ correlators in the topological side. This relation was pointed out by authors in Ref.~\cite{YixinZhang}. For completeness, we derive it here to make the discussion self-contained.

The model contains charge-$1e$ spinless fermions and charge-$2e$ bosons representing $p$-wave Cooper pairs. If both the fermions and bosons are dilute, we can take the long-wavelength continuum limit and focus on the low-energy degrees of freedom near the band minima. Furthermore, the continuum limit can be taken only for bosonic degrees of freedom, if the superconducting phase fluctuates slowly in time and space \cite{physrevb.98.214501} and the fermionic sector has a gap. 

The two sectors are coupled through a charge-conserving pairing term, which allows a Cooper pair to be converted into a pair of fermions and vice versa. Microscopically, the coupling is given by
\begin{equation}
        H_{cb}=u\sum_{i}c_i^\dagger c_{i+1}^\dagger e^{2i\phi_i}+\text{H.c.}
\end{equation}
In the continuum limit, this becomes
\begin{equation}
        H_{cb}=u\int dxe^{2i\phi_x}c^\dagger_x \partial_x c^\dagger_x +\text{H.c.}
\end{equation}
Here, $e^{2i\phi}$ annihilates a charge-$2e$ Cooper pair.

In one dimension, a configuration of bosons is completely specified by their density distribution,
\begin{equation}
        \rho(x)=\sum_{i}\delta(x-x_i)
\end{equation}
which records the position $x_i$ of each particle.

The starting point of the bosonization analysis \cite{Giamarchi2003} is to introduce a phase field $\theta(x)$ that encodes the particle positions according to
\begin{equation}
        \theta(x_i)=2\pi i,i\in\mathbb{Z},\frac{d\theta}{dx}>0
\end{equation}
We separate the smooth density fluctuations from the uniform background by writing the particle-position field as $\theta(x)\to\theta(x)+2\pi\bar{\rho}_bx$, where $\bar{\rho}_b$ is the average boson density. The density can then be expressed as
\begin{equation}
        \begin{split}
        \rho(x)&=\sum_i\delta(\theta(x)+2\pi\bar{\rho}_bx-2\pi i)\left(\frac{d\theta}{dx}+2\pi\bar{\rho}_b\right)\\
                &=\sum_i\int\frac{dk}{2\pi}e^{ik(\theta(x)+2\pi\bar{\rho}_bx-2\pi i)}\left(\frac{d\theta}{dx}+2\pi\bar{\rho}_b\right)\\
                &=\sum_{n\in\mathbb{Z}} \exp(in(\theta(x)+2\pi\bar{\rho}_bx))\left(\frac{\partial\theta}{2\pi}+\bar{\rho}_b\right)\\
                &\approx \frac{\partial\theta}{2\pi}+\bar{\rho}_b + \sum_{n\neq 0}\exp(in(\theta(x)+2\pi\bar{\rho}_bx))
        \end{split}
\end{equation}
where the last line retains the leading slowly varying contribution from the prefactor. Thus, the smooth density fluctuation is related to $\theta$ by $\delta\rho(x)=\frac{\partial_x\theta}{2\pi}$. The oscillarting terms describe density-wave fluctuations. In particular, the factor $e^{i2\pi n\bar{\rho}_bx}$ shows that the operator $e^{in\theta}$ carries momentum $2\pi n\bar{\rho}_b$.

The field $\phi$ dual to $\theta$ is defined by the commutation relation
\begin{equation}
        [\phi_y,\theta_x]=i\pi\Theta(y-x),e^{2i\phi_y}\rho_xe^{-2i\phi_y}=\rho_x+2i[\phi_y,\frac{\partial\theta_x}{2\pi}]=\rho_x+\delta(x-y)
\end{equation}
Thus, $e^{2i\phi}$ annihilates a boson, or equivalently a charge-$2e$ Cooper pair in the present model, consistent with identification made above. Moreover, $e^{-i\theta_x}$ can create a $\pi$-kink in $\phi_y$.  

The low-energy effective Hamiltonian density of boson is
\begin{equation}
    \mathcal{H}_b=\frac{\rho_s}{2}(\partial 2\phi)^2+\frac{1}{2\chi}(\frac{\partial \theta}{2\pi})^2-2\mu(\bar{\rho}_b+\frac{\partial\theta}{2\pi})
\end{equation} 
Here, $\rho_s$ and $\chi$ denote the superfluid stiffness and compressibility of the bosons, respectively. 

The fermionic sector in the continuum limit is described by
\begin{equation}
    \mathcal{H}_c=c^\dagger (\epsilon_0-\frac{\partial^2}{2m}-\mu)c +u(c^\dagger \partial c^\dagger e^{2i\phi}+\text{H.c.})
\end{equation}
The electron field has been expanded around a single momentum at the band bottom. The derivative in the pairing term gives the characteristic $p$-wave pairing structure.

The local operators can be classified according to their charge $Q$ and momentum $P$ quantum numbers. For example, the opeators with $(Q,P)=(1,0)$ include $c^\dagger,ce^{-2i\phi}$, while the operator with $(Q,P)=(0,1)$ is $e^{i\theta}$. To partially decouple the bosonic and fermionic sectors, we perform the unitary transformation
\begin{equation}
    U=\exp\left(-i\int dx \phi(x)(c_x^\dagger c_x-\bar{\rho}_c)\right)
\end{equation}
Under this transformation, the local operators become
\begin{equation}
        Uc_xU^{-1}=f_x e^{i\phi(x)},Ue^{i\theta(x)}U^{-1}=e^{i\theta(x)}\exp(i\pi\int_x^{\infty}dy(f_y^\dagger f_y-\bar{\rho}_c))
\end{equation}
Here, $f$ denotes the neutral fermion operator after the transformation. The transformation introduces a $\mathbb{Z}_2$ redundancy in the decomposition of the physical fermion, $f_x\to-f_x,e^{i\phi(x)}\to-e^{i\phi(x)}$. This redundancy does not affect physical observables, since all local physical operators are invariant under this transformation. In particular, although $e^{i\phi}$ by itself is not a well-defined local physical operator, it always appears in combinations with $f$ that are invariant under $\mathbb{Z}_2$ transformation.
Importantly, the finite-momentum boson density operator $e^{i\theta_x}$ is dressed by a nonlocal string of the neutral fermion density. 

After the transformation, the Hamiltonian density takes the form
\begin{equation}
        \begin{split}
        \mathcal{H}_b+\mathcal{H}_f&=f^\dagger (\epsilon_0-\frac{\partial^2}{2m})f+u(f^\dagger \partial f^\dagger +\text{H.c.})\\
        &+\frac{\rho_s}{2}(\partial 2\phi)^2+\frac{1}{2\chi}(\frac{\partial \theta}{2\pi})^2\\
        &-\mu(2\bar{\rho}_b+\bar{\rho}_c+\frac{\partial\theta}{\pi})
        \end{split}
\end{equation}
Here, the residual coupling between fermionic and bosonic sectors has been omitted, as it is irrelevant in the low-energy phases considered below \cite{Kane2017}.

The chemical potential term therefore identifies the long-wavelength physical charge density $\rho_e$ as $2\bar{\rho}_b+\bar{\rho}_c+\frac{\partial\theta}{\pi}$. The Hamiltonian of $f$ fermions is the continuum limit of the Kitaev-chain model near its topological phase transition. The $\epsilon_0<0$ regime corresponds to the topologically nontrivial phase, while $\epsilon_0>0$ corresponds to the topologically trivial phase. Away from the critical point $\epsilon_0=0$, the fermion sector is gapped. Consequently, the coupling between the fermionic and bosonic sector is irrelevant at low energies and can be neglected in the phases considered below. 

The Kitaev-chain admits a dual description in terms of the one-dimensional transverse-field Ising model.
\begin{equation}
    H=-J\sum_j\sigma_j^x\sigma_{j+1}^x-h\sum_j\sigma_j^z
\end{equation}
The phases of the Ising model can be distinguished by the long-distance behavior of the correlation function of the $\mathbb{Z}_2$ order parameters $\langle \sigma_i^x\sigma_j^x\rangle$.
In the $\mathbb{Z}_2$ symmetry-broken phase, $J>h$, this correlation approaches to a nonzero constant in the long-distance limit, while correlations of the transverse operators $\sigma_z,\sigma_y$ decay exponentially at long distances.
Considering the Jordan-Wigner transformation
\begin{equation}
    \begin{split}
\frac{\sigma_j^x+i\sigma_j^y}{2}&=f_j (-1)^{\sum_{i>j}n_i^f}\\
\frac{\sigma_j^x-i\sigma_j^y}{2}&=f_j^\dagger (-1)^{\sum_{i>j}n_i^f}\\
\sigma_j^z&=1-2n_j^f
    \end{split}
\end{equation}
the Ising model is mapped onto the Kitaev chain,
\begin{equation}
        H=-J\sum_j\left(f_{j+1}^\dagger f_j+f_{j}^\dagger f_{j+1}-f_j^\dagger f_{j+1}^\dagger -f_{j+1}f_j\right)+2h\sum_jf_j^\dagger f_j
\end{equation}
The exactly solvable Kitaev chain is a topological superconductor for $0<h<J$, characterized by Majorana zero modes localized at its boundaries. For $h>J$, it is in the topologically trivial superconducting phase. 

The correspondence is directly reflected on the Ising-model side. For $0<h<J$, the Ising model is in the spontaneous $\mathbb{Z}_2$-symmetry-broken phase, whereas for $h>J$ it is in the paramagnetic phase. Thus, the ordered phase of the Ising model corresponds to the topological superconducting phase of the Kitaev chain, while the paramagnetic phase corresponds to its topologically trivial phase. 

Considering the nonlocal string operator in the Kitaev chain,
\begin{equation}
        \langle f_i(-1)^{\sum_{i<k<j}n_k^f}f_j^\dagger \rangle=-\langle \sigma^-_i\sigma^+_j\rangle\sim \langle f_x\exp(i\pi\int_x^yds n_s^f)f_y^\dagger\rangle
\end{equation}
In the topological superconducting phase, this nonlocal string correlator exhibits long-range order. This follows from the duality to the $\mathbb{Z}_2$ symmetry-broken phase of the Ising model, where the string operator is mapped onto the correlations of $\mathbb{Z}_2$ order parameter. 

Motivated by this correspondence, we now consider the analogous Jordan-Wigner string operator in the coupled fermion-boson model. After the unitary transformation, the string generated by the physical charge density is 
\begin{equation}
        (-1)^{\sum_{i<k<j}(c_i^\dagger c_i+2n^b_i)}\to e^{i2\pi\bar{\rho}_b(y-x)}e^{i\pi\bar{\rho}_c(y-x)}e^{i\theta(y)}e^{-i\theta(x)}
\end{equation}
At the endpoints of the string, we attach local operators carrying charge and momentum $(Q,P)=(1,n)$. Microscopically, this corresponds to measuring single-particle $(Q=1)$ correlators connected by a Jordan-Wigner string. For a given momentum $n$, the corresponding operator in the correlation is
\begin{equation}
        f_xe^{i\phi(x)}e^{in\theta(x)}\exp(in\pi\int_x^ydsf^\dagger_sf_s)e^{-in\pi\bar{\rho}_c(y-x)}e^{i2\pi n\bar{\rho}_b(x-y)}f_y^\dagger e^{-in\theta(y)}e^{-i\phi(y)}
\end{equation}
Combining this operator with the Jordan-Wigner string generated by the charge density, we obtain the nonlocal correlation function which is measured as $A(r)$ in DMRG simulations
\begin{equation}
        A(x-y)\sim\sum_{n\in\mathbb{Z}}\langle f_x e^{i\phi(x)}e^{i(n-1)\theta(x)}\exp(in\pi\int_x^ydsf_s^\dagger f_s)e^{i\pi(n-1)\bar{\rho}_e(x-y)}e^{-i\phi(y)}e^{-i(n-1)\theta(y)}f_y^\dagger \rangle
\end{equation}
In the topological superconducting phase, the nonlocal string correlation in the $f$-fermion sector exhibits long-range order for odd $n$. The leading contribution corresponds to $n=1$, for which the correlator decays as $r^{-K_{\text{string}}}$. Meanwhile, the Cooper-pair correlator associated with the operator $e^{2i\phi}$ exhibits power-law decay, $r^{-K_{\text{SC}}}$. Since the $\phi$ field is described by a quadratic low-energy Hamiltonian, the scaling exponents of these correlators are related by
\begin{equation}
        K_{\text{SC}}=4K_{\text{string}}
\end{equation}
In the topologically trivial phase, the Jordan–Wigner-connected single-particle correlation function is always short-ranged. For odd $n$, the nonlocal string correlation in $f$-sector is short-ranged because the corresponding Ising sector is in the disordered phase. For even $n$, the fermionic string does not acquire long-range order either, and the correlation function remains short-ranged due to the gap in the $f$-secotr. Thus, irrespective of the momentum $n$, the Jordan-Wigner-connected single-particle correlation function decays exponentially in the topologically trivial phase. 

The single-particle operators carrying non-zero momentum are  
\begin{equation}
        \begin{split}
       e^{i2\pi\bar{\rho}_bx} Uc_x e^{i\theta_x}U^{-1}&=f_x e^{i\phi_x+i\theta_x+i\pi\int_x^\infty ds(f_s^\dagger f_s-\bar{\rho}_c)+i2\pi\bar{\rho}_bx}\\
        e^{-i2\pi\bar{\rho}_bx} Uc_x e^{-i\theta_x}U^{-1}&=f_x e^{i\phi_x-i\theta_x-i\pi\int_x^\infty ds(f_s^\dagger f_s-\bar{\rho}_c)-i2\pi\bar{\rho}_bx}
        \end{split}
\end{equation}
The corresponding correlation functions are
\begin{equation}
        \begin{split}
        e^{i2\pi\bar{\rho}_b(x-y)}\langle c_x e^{i\theta_x}e^{-i\theta_y}c_y^\dagger\rangle &= \langle f_xe^{i\pi\int_x^ydsf_s^\dagger f_s}f_y^\dagger e^{i(\phi_x+\theta_x)} e^{-i(\phi_y+\theta_y)} \rangle e^{ik_f (x-y)}\\
        e^{-i2\pi\bar{\rho}_b(x-y)}\langle c_x e^{-i\theta_x}e^{i\theta_y}c_y^\dagger\rangle &= \langle f_xe^{i\pi\int_x^ydsf_s^\dagger f_s}f_y^\dagger e^{i(\phi_x-\theta_x)}e^{i(\phi_y-\theta_y)} \rangle e^{-ik_f (x-y)}
        \end{split}
\end{equation}
where $k_f=\pi(2\bar{\rho}_b+\bar{\rho}_c)$. The momentum of the single-particle operator is fixed by the total charge density. According to the Luttinger theorem, the charge density determines the Fermi momentum of the low-energy Luttinger liquid. We therefore identify $c_x e^{i\theta_x}e^{i2\pi\bar{\rho}_bx}$ and $c_x e^{-i\theta_x}e^{-i2\pi\bar{\rho}_bx}$ with the right- and left-moving components of the electron opeator, respectively. 

In the topological phase of the Kitaev model, the single-particle correlation $G(r)$ is reduced to 
\begin{equation}
    G(x-y)\sim e^{i(\theta_x+\phi_x)}e^{-i(\theta_y+\phi_y)} e^{ik_f(x-y)}+e^{-i(\theta_x-\phi_x)}e^{i(\theta_y-\phi_y)} e^{-ik_f(x-y)}
\end{equation}
Thus, the single-particle excitations are gapless due to the power-law correlations of vertex operators $e^{i(\phi_x\pm\theta_x)}$. In the topologically trivial phase, $G(r)$ decays exponentially as nonlocal correlations of $f$ are short-ranged in the topologically trivial side of the Kitaev chain.

The $2k_f$-density modulation is $\rho_{2k_f}(x)\sim e^{2i\theta_x}$. Its correlation in the unitary transformed theory is
\begin{equation}
        \langle \rho_{2k_f}(x)\rho_{2k_f}(y)^\dagger \rangle =\langle e^{2i\theta_x}e^{-2i\theta_y}\rangle 
\end{equation}
For the original local operator $e^{i\theta_x}$, the correlation function can be written as 
\begin{equation}
        \langle e^{i\theta_x}e^{-i\theta_y}\rangle e^{i2\pi\bar{\rho}_b(x-y)}=\langle e^{i\theta_x}e^{-i\theta_y}\exp(i\pi\int_x^ydsf_s^\dagger f_s)\rangle e^{ik_f(x-y)}
\end{equation}
The string operator is mapped onto a nonlocal operator in the Ising model side
\begin{equation}
        \langle \exp(i\pi\int_x^yds f_s^\dagger f_s)\rangle=\langle \prod_{i<k<j}\sigma_k^z\rangle
\end{equation}
This operator can be investigated by employing the Kramers-Wannier duality.
\begin{equation}
        \sigma_j^z=\tau_{j-\frac{1}{2}}^x\tau_{j+\frac{1}{2}}^x,\sigma_j^x\sigma_{j+1}^x =\tau_{j+\frac{1}{2}}^z 
\end{equation}
Consequently, the fermion-parity string is mapped onto a correlation function of the dual Ising spins $\langle \tau_i^x\tau_j^x\rangle$. The Ising model becomes
\begin{equation}
        H=-J\sum_j\tau_{j+\frac{1}{2}}^z-h\sum_j\tau_{j-\frac{1}{2}}^x\tau_{j+\frac{1}{2}}^x
\end{equation}
In the topological phase, $J>h$, the dual Ising model is in its disordered phase, and the correlation function $\langle \tau_i^x\tau_j^x\rangle$ is therefore short-ranged. Hence, the pure fermion-parity string does not contribute a long-range factor to the correlation function $\langle e^{i\theta_x}e^{-i\theta_y}\rangle$. This indicates that the Cooper-pair density-wave correlation does not exhibit quasi-long-range order.

\end{document}